\documentclass[a4paper,11pt]{article}
\usepackage{jinstpub} 
\usepackage{lineno}

\usepackage{amsmath} 
\usepackage{euscript}
\usepackage{hyperref} 
\hypersetup{
    colorlinks=true,
    linkcolor=blue,
    filecolor=magenta,  
    citecolor=blue,
    urlcolor=blue 
    }
\usepackage[all]{hypcap} 
\usepackage[compat=1.1.0]{tikz-feynman}

\title{\boldmath Domain-Adversarial Graph Neural Networks for $\Lambda$ Hyperon Identification with CLAS12}

\author[a,1]{M. McEneaney,\note{Corresponding author.}}
\author[a,b]{A. Vossen}
\affiliation[a]{Department of Physics, Duke University,\\120 Science Drive, Durham, NC 27708, USA}
\affiliation[b]{Thomas Jefferson National Accelerator Facility,\\12000 Jefferson Ave., Newport News, VA 23606, USA}

\emailAdd{matthew.mceneaney@duke.edu}

\abstract{
Machine learning methods and in particular Graph Neural Networks (GNNs) have revolutionized many tasks within the high energy physics community.  Particularly in the realm of jet tagging, GNNs and domain adaptation have been especially successful.  However, applications with lower energy events have not received as much attention.  We report on the novel use of GNNs and a domain-adversarial training method to identify $\Lambda$ hyperon events with the CLAS12 experiment at Jefferson Lab.  The GNN method we have developed increases the purity of the $\Lambda$ yield by a factor of $1.95$ and by $1.82$ using the domain-adversarial training.  This work also provides a good benchmark for developing event tagging machine learning methods for the $\Lambda$ and other channels at CLAS12 and other experiments, such as the planned Electron Ion Collider.
}


\arxivnumber{2302.05481} 

\begin{document}
\maketitle
\flushbottom

\section{Introduction} \label{Introduction}
The internal dynamics of the nucleon are governed by the strong interaction, one of the four fundamental forces in nature, and present a significant and yet unconquered intellectual challenge.  As the name suggests, the strong interaction creates a strongly coupled system within the nucleon making it extremely difficult to calculate.  Fortunately, due to the asymptotic freedom of the strong interaction at high energies, one way to study these interactions is through scattering events where subatomic particles such as electrons and protons are accelerated to high energies and collided together.  The momentum and trajectories of particles produced in these interactions may be measured with a variety of detector systems.  From this final state information, one may begin to infer the original dynamics of the nucleon before the collision.

The $\Lambda$ hyperon is typically detected via the two-body decay $\Lambda \rightarrow p \pi^{-}$.  It is of particular interest because the polarization of the $\Lambda$ is preserved in the cross-section of the decay protons, allowing one to infer information about the spin structure of the $\Lambda$.  One may identify the $\Lambda$ signal by looking at the invariant mass spectrum of the proton pion ($p\pi^{-}$) pairs for a peak around the nominal $\Lambda$ mass of $1.1157$~GeV.  However, at CLAS12 this becomes complicated because of the large background from non-strange final states and combinatorics.  One of the goals of this study has been to develop a method of increasing the $\Lambda$ signal to background ratio while still preserving the signal shape in order to be able to extract the $\Lambda$ yield from a fit to the signal.


Graph Neural Networks (GNNs) have been used to great efficacy for many different applications in particle physics~\cite{thais2022graph}.  In particular, event-level classification has seen many successful applications of GNNs, for example at the Large Hadron Collider (LHC) tagging jet types using the four-momenta of the constituent particles~\cite{PhysRevD.101.056019}.  Domain adaptation methods have also proven quite successful in mitigating differences in neural network performance between training and real data~\citep{CMS_LLP_2020,clavijo2021adversarial}.  However, the application of GNNs and Domain Adaptation in the realm of tagging resonances in lower energy SIDIS events has not received as much attention, to the authors' knowledge.

GNNs build on the concept of a neural network by treating each node of the input graph as an input channel of a neural network layer. Node values in subsequent layers are obtained by computing a convolution over neighboring nodes within the current layer.  GNNs are well-suited to event-level classification tasks because they are permutation invariant with respect to the order of the input particles and they can handle a variably sized input.  GNNs also rely on the correlations between the nodes of the input graph which makes them a natural choice for identifying events based on the products of a decay.

\subsection{Physics Observables} \label{Physics Observables}
\begin{figure}[h!]
\centering
\begin{tikzpicture}
  \begin{feynman}


     \vertex (a) at (0,0) {$\ell$}; 
    \vertex (b) at (1,0);
    \vertex (c) at (2,1) {$\ell'$};
    \vertex (d) at (2,-1);
    \node [circle,draw=black,thick] (e) at (1,-2) {$N$};
    \vertex (f) at (3,-1) {CFR};
    \vertex (g) at (3,-2) {TFR};

    \diagram* {
      (a) -- [fermion] (b) -- [fermion] (c),
      (b) -- [boson, edge label'=\(\gamma^{*}\)] (d),
      (e) -- [fermion, edge label'=\(q\)] (d),
      (d) -- [fermion, edge label'=\(q'\)] (f),
      (e) -- [thick] (g),
    };
  \end{feynman}
\end{tikzpicture}
\caption{Semi-Inclusive Deep Inelastic Scattering}
\label{fig:sidis}
\end{figure}
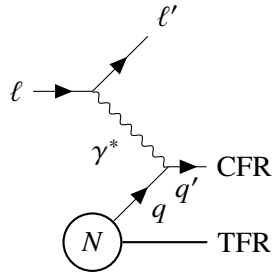

Semi-Inclusive Deep Inelastic Scattering (SIDIS)~\cite{Aidala_2013} events occur when a high-energy lepton $\ell$ collides with a target nucleon $N$ and the scattered lepton and at least one hadron are detected in the final state.  SIDIS provides one of the ideal laboratories in which to probe the internal dynamics of the nucleon.  The hadrons may be produced from the struck quark $q'$, i.e. in the Current Fragmentation Region (CFR), or from the remnant quarks, i.e. in the Target Fragmentation Region (TFR) as shown in figure~\ref{fig:sidis}.  Characteristics of final-state hadrons and correlations between them can reveal information about the spin and momentum structure of the nucleon.  Some relevant kinematic variables for this study include the following:

\begin{equation}
    Q^2 = -q^2
\label{eq:Q2}
\end{equation}
\begin{equation}
    W^2 = (P+q)^2
\label{eq:W2}
\end{equation}
\begin{equation}
    \nu = E - E'
\label{eq:nu}
\end{equation}
\begin{equation}
    x = \frac{Q^2}{2 P \cdot q}
\label{eq:x}
\end{equation}
\begin{equation}
    y = \frac{P \cdot q}{P \cdot \ell}
\label{eq:y}
\end{equation}
\begin{equation}
    z = \frac{P \cdot P_h}{P \cdot q}
\label{eq:z}
\end{equation}
\begin{equation}
    x_F = \frac{2 P_h \cdot q}{W \lvert q \rvert}.
\label{eq:xF}
\end{equation}

The virtual photon $\gamma^{*}$ from the interaction of the incoming lepton $\ell$ with the struck quark imparts a momentum transfer $\ell-\ell'$ with virtuality $Q^2$.  The incoming lepton has energy $E$ and the outgoing lepton $\ell'$ has energy $E'$ such that $\nu$ is the difference in energy between $\ell$ and $\ell'$ and $y$ is the fraction of the incoming lepton energy transferred to the virtual photon.  The total energy in the $\gamma^{*}N$ center of mass frame is $W$ and the fraction of the nucleon momentum carried by the struck quark is $x$.  A final-state hadron $h$ carries a fraction $z$ of the virtual photon energy and a fraction $x_F$, the so-called $x$-Feynman variable, of the hadron's longitudinal momentum in the virtual photon direction relative to the maximum value $W/2$ allowed by momentum conservation in the struck-quark center of mass frame.  $x_F$ is used to distinguish between the CFR and TFR since it favors positive values for hadrons travelling in the direction of the incoming virtual photon and negative values for backward-going hadrons.

\subsection{The CLAS12 Experiment} \label{The CLAS12 Experiment}
In order to realize these high energy scattering events the Continuous Electron Beam Accelerator Facility (CEBAF) at Jefferson Lab delivers a high luminosity polarized electron beam to four experimental halls for fixed target experiments~\cite{doi:10.1146/annurev.nucl.51.101701.132327}.  The CLAS12 (CEBAF Large Acceptance Spectrometer $12$~GeV) Experiment is located in the experimental Hall B and provides excellent momentum and angular coverage and good particle identification capabilities for both charged and neutral particles produced in high energy electron-proton or electron-deuteron scattering events~\cite{BURKERT2020163419}.  The detector is centered around two large magnets, a solenoid in the central region of the detector and a torus magnet in the forward region.  The torus magnet is operated in two different configurations, either bending negatively charged particles in toward the beamline (inbending configuration), or out away from the beamline (outbending configuration).

\section{Data} \label{Data}
The data used in this study was all taken during the fall 2018 run period in the outbending toroidal configuration with a 10.6~GeV polarized electron beam and an unpolarized liquid hydrogen target.  Events were required to have an identified  scattered electron ($e^{-}$) and a proton-pion ($p\pi^{-}$) pair in the reconstruction.  Further restrictions on the scattered electron were that it be the trigger particle, be detected in the forward region of the detector, and have the highest momentum of all electrons in the event.  The scattered electron was also required to have a particle identification (PID) assignment quality factor estimate $|\chi^2|<3$.  Note that the $\chi^2$ PID quality factor is not an actual $\chi^2$ statistic value although it has the same name.  For electrons it is related to the difference of the energy deposition from an expected value estimated from the measured momentum, and for charged hadrons it is related to the difference from the expected vertex time.  No estimate is given for neutral particles.  Kinematic cuts were $Q^2>1$~GeV$^2$, $W>2$~GeV, $y<0.8$, $z_{p\pi^{-}}<1$ and $x_F>0$ for the $p\pi^{-}$ pair to select $\Lambda$s produced in the CFR.  Additionally, the invariant mass of the $p\pi^{-}$ pair was restricted to $M_{p\pi^{-}}<1.24$~GeV so that events were sufficiently close to the $\Lambda$ mass peak.

\subsection{Monte Carlo Data Set} \label{Monte Carlo Data Set}
\begin{figure}[h!]
\centering
\includegraphics[width=0.45\linewidth]{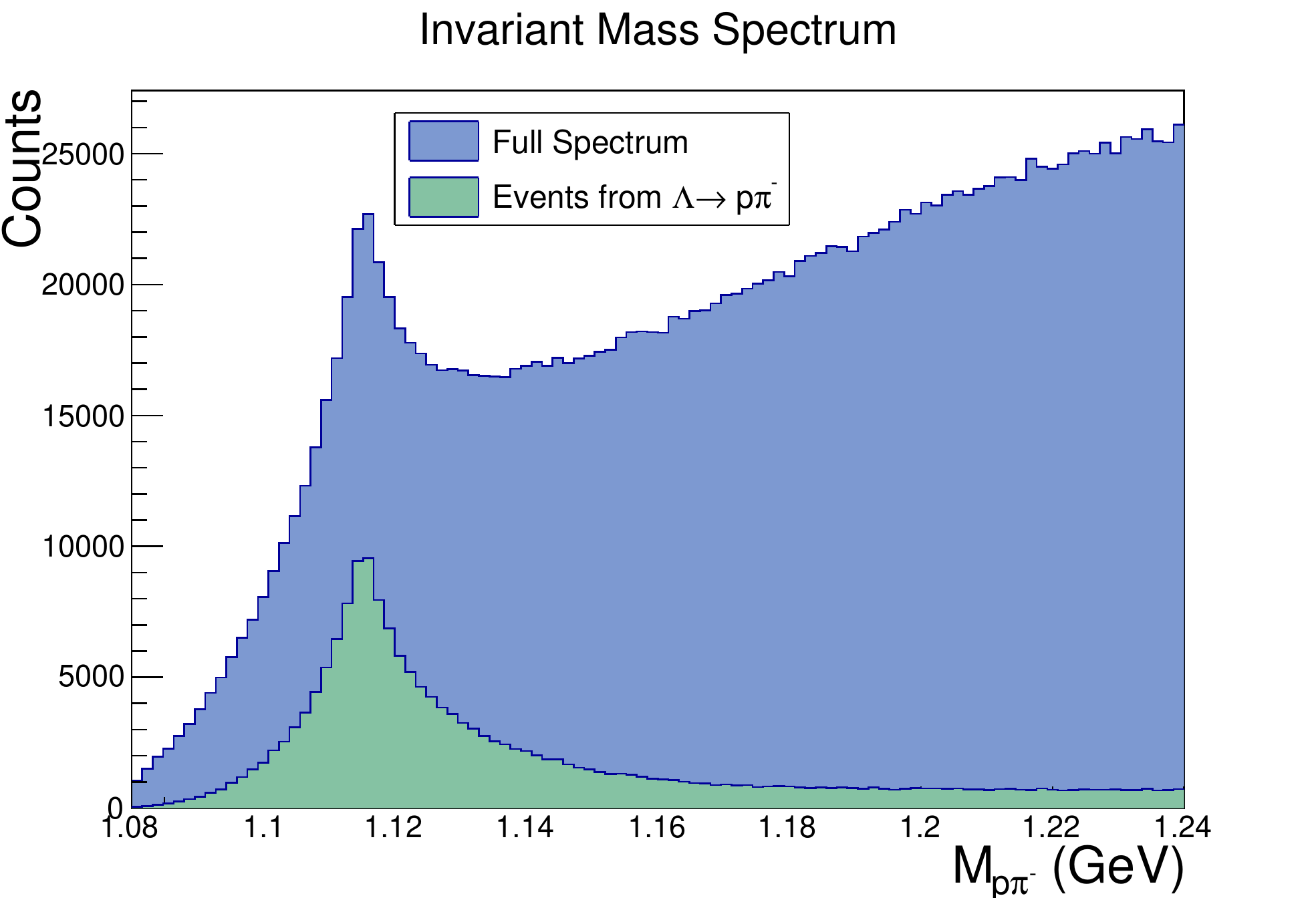}
\caption{The MC invariant mass spectrum for reconstructed $p\pi^{-}$ pairs is shown for both the full set of events and signal events with a $\Lambda$ decay in the MC truth.}
\label{fig:truthmatching}
\end{figure}

The sample of Monte Carlo (MC) simulation events was produced with the same run configuration as the actual data from the fall 2018 run period.  Events were generated with an MC algorithm based on the Pepsi Lund program~\cite{MANKIEWICZ1992305}.  For the purpose of training our neural networks, $\Lambda$ signal events were identified as those that contained a $\Lambda \rightarrow p\pi^{-}$ decay in the MC truth.  A comparison of the full invariant mass spectrum to the spectrum of signal events is shown in figure~\ref{fig:truthmatching}.

\subsection{Signal Extraction} \label{Signal Extraction}
\begin{figure}[h!]
\centering
\includegraphics[width=0.45\textwidth]{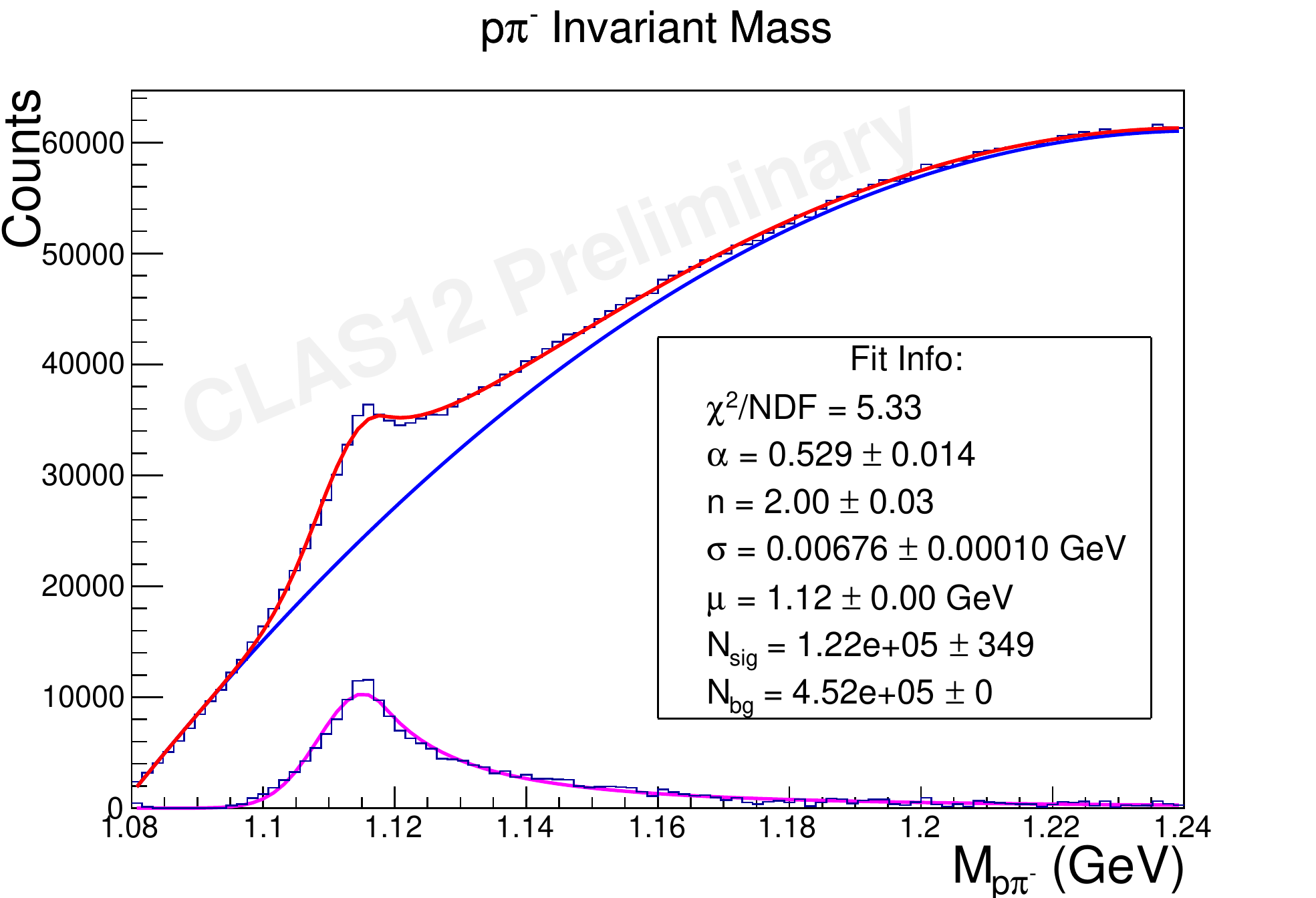}
\includegraphics[width=0.45\textwidth]{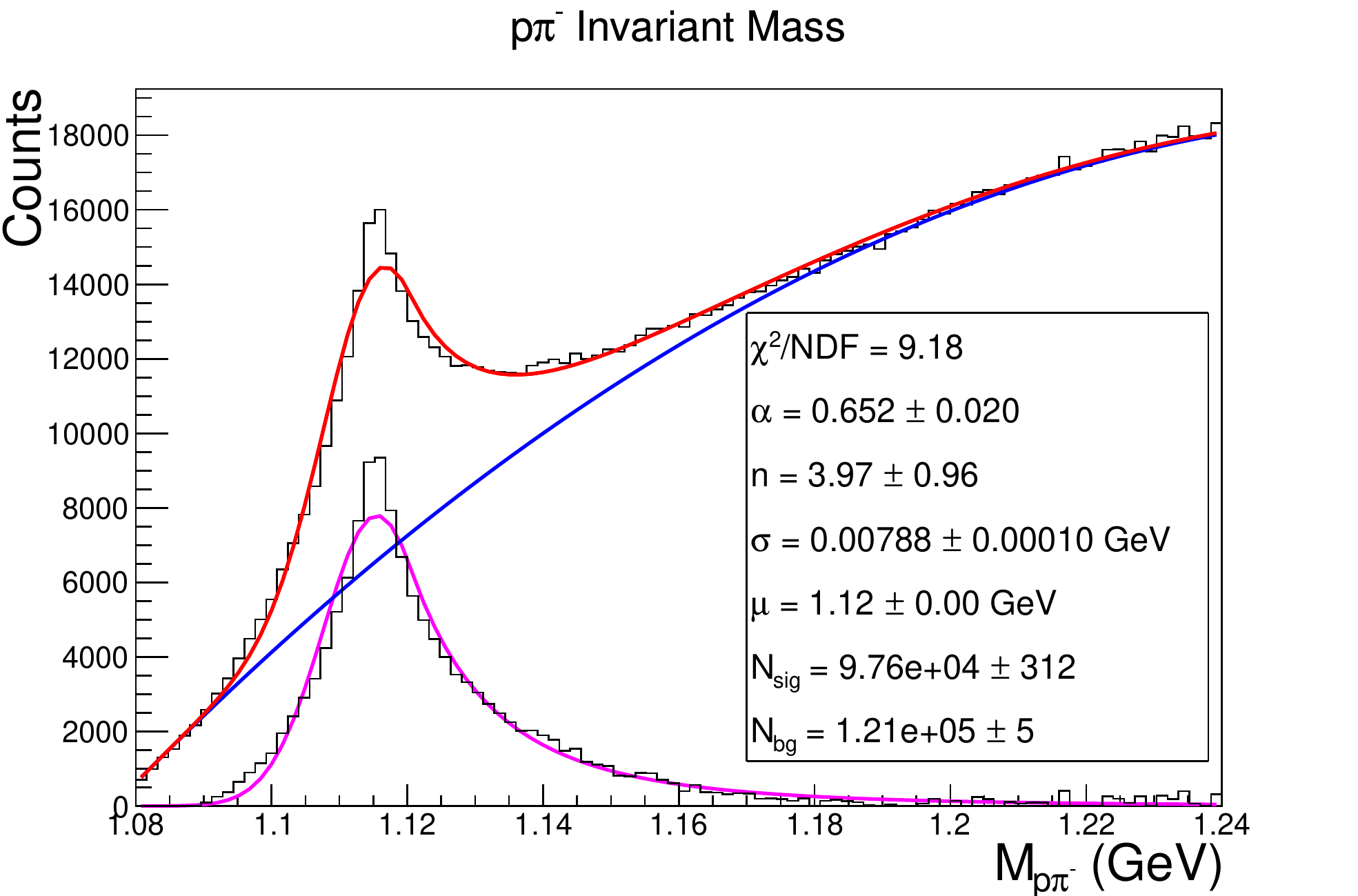}
\caption{A Crystal Ball signal over a quadratic background fit function was used to extract the $\Lambda$ signal statistics from data (left) and MC (right). The number of signal events $N_{sig}$ is estimated using the histogram counts over the fitted background function (small outlined histogram).}
\label{fig:mass} 
\end{figure}

The invariant mass spectrum of all reconstructed $p\pi^{-}$ pairs passing kinematic cuts is shown in figure~\ref{fig:mass} for both the real data and MC simulation.  A peak around the nominal $\Lambda$ mass $M=1.1157$~GeV is apparent, but the background contribution is very high especially in data.  The main background contributions come from the $\Delta^{0}\rightarrow p\pi^{-}$ decay which peaks at $1.232$~GeV and from combinatorics of random $p\pi^{-}$ pairs.  A Crystal Ball~\cite{PhysRevD.34.711} signal over a quadratic background was used to model the invariant mass spectrum.  The Crystal Ball function consists of a Gaussian peak with a power-law tail such that both the function and its first derivative are continuous
\begin{eqnarray}
    f(x;\alpha,n,\mu,\sigma) = 
    N
    \left\{
    \begin{array}{lr}
        \exp{-\frac{(x-\mu)^{2}}{2\sigma^{2}}}, & \frac{x-\mu}{\sigma} > \alpha\\\\
        A (B - \frac{x-\mu}{\sigma})^{-n}, & \frac{x-\mu}{\sigma} \leq \alpha
    \end{array}
    \right.
    \label{eq:crystalball}
\end{eqnarray}
where $N$, $A$, and $B$ are constants defined by the function parameters and satisfying the continuity requirements and normalization.  The signal is much more pronounced in MC but the fit parameters were similar up to scale factors between MC and data.  The number of signal events $N_{sig}$ is estimated using the histogram counts over the fitted background function within a $\pm2\sigma$ region around the signal mean $\mu$ from the fit.  The histogram counts are used instead of the fit function integral since the histogram has some noticeable excess above the fit function at the signal peak.  The signal fit region limits have been optimized so as to maximize the signal purity and minimize the signal error.

\section{Machine Learning Methods} \label{Machine Learning Methods}
The following sections assume a basic familiarity with supervised learning and neural networks; however, a comprehensive introduction may be found in ref.~\cite{DeepLearning}.

\subsection{Graph Neural Networks} \label{Graph Neural Networks}
\begin{figure*}[h!]
\centering
\includegraphics[width=0.90\textwidth]{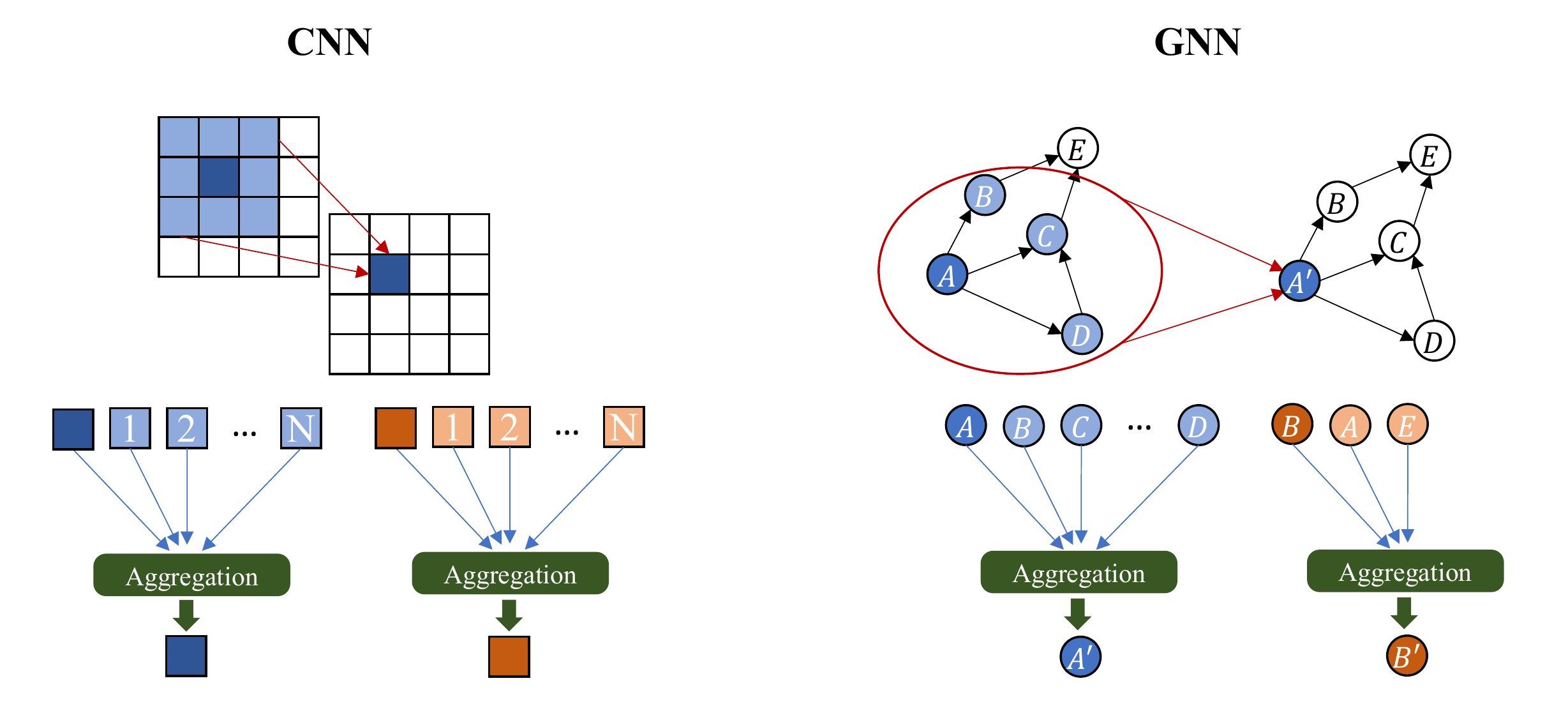}
\caption{Convolution in a GNN is analogous to the convolution step in a CNN except that the neighbors of each node are defined by the structure of the graph.}
\label{fig:cnn_vs_gnn}
\end{figure*}

GNNs are well suited to many physics applications since they are able to take inputs with irregular and unordered structure, e.g. unordered sets of varying numbers of final-state particles in a SIDIS event.  A graph may be described as a set of vertices $V$ and edges $E$.  At the simplest level, GNNs operate similarly to Convolutional Neural Nets (CNNs) by taking a convolution over the neighborhood of each node to compute its representation in the following layer as depicted in figure~\ref{fig:cnn_vs_gnn}.  CNNs take a rectangular input such as a two-dimensional pixel array and use an aggregation of the information from the neighboring pixels to compute the representation of a pixel in the next layer.  GNN convolution works the same way except that the neighbors are determined by the structure of the input graph, which may be highly irregular.  In fact, CNNs are a subclass of GNNs since a rectangular array is in effect just a graph with a regular rectangular structure.

Specifically, to update the graph representation in the $k^{th}$ layer of a network, the representation $h_m^{(k)}$ of a given node or edge $m$  is formed from some aggregation function $A$ of the information contained in the neighboring nodes and edges in the previous layer.

\begin{equation}
    \label{node_update}
    h_m^{(k)} = A( \{h_n^{(k-1)} : n \in \EuScript{N} \} ).
\end{equation}
The aggregation function must not depend on the order of the inputs and must be able to handle a variable number of arguments, e.g. a sum, maximum, or minimum function.  The set of neighbors $\EuScript{N}$ of a node $m$ is defined by the structure of the input graph and may include the node $m$ itself.  A more detailed discussion may be found in ref.~\cite{https://doi.org/10.48550/arxiv.1806.01261}.  Using this concept of graph convolution, many different types of convolutional blocks may be formed, and these blocks may be combined to construct the different layers of a GNN.

\subsubsection{Graph Isomorphism Networks} \label{Graph Isomorphism Networks}
Graph Isomorphism Networks (GINs) are theoretically the most powerful form of GNNs and have been quite successful at graph classification tasks~\cite{xu2019powerful}.  GINs essentially add two important modifications to the graph convolution algorithm.  After the aggregation step over the neighboring nodes in the graph, a Multi-Layer Perceptron (MLP) is introduced so that on the $k^{th}$ iteration the updated node $h^{(k)}$ becomes

\begin{equation}
    \label{node_update_GIN}
    h_m^{(k)} = MLP\big{(} (1+\epsilon) \cdot h_m^{(k-1)} + A( \{h_n^{(k-1)} : n \in \EuScript{N}\} ) \big{)},
\end{equation} 
where $\epsilon$ is an optional learnable parameter and $\EuScript{N}$ is the set of all neighbors of node $h_m$~\cite{xu2019powerful}.

The second modification is that in the final layer the hidden representations of the graph from each layer of the GIN are pooled together to obtain the final representation.  The reasoning for this type of pooling is that localized information can be lost deeper in the network.  Pooling over all previous representations gives a good compromise between the localized information contained within the initial layers and more generalized information reached in later ones.

\subsection{Domain Adversarial Networks} \label{Domain Adversarial Networks} 
Supervised learning relies on the training sample being representative of actual data.  With SIDIS events, this means the simulation events should closely match the real data events intended for classification.  However, simulation cannot perfectly match data for all physics channels.  Thus, performance on real data may not live up to performance on MC test events.

One way to overcome this without trying to completely re-tune the simulation is to use a domain-adversarial training routine first introduced in~\cite{https://doi.org/10.48550/arxiv.1505.07818}.  If there is some shared phase space between the training sample and the real data another training objective may be introduced.  In addition to minimizing the loss of the classifier, one also maximizes the loss of an additional head network on top of the base network (in this case the GNN) which is simultaneously trained to distinguish between training data and data from the real test domain.  This forces the base network to converge towards a hidden representation that relies only on features common to both training and test domains.

\begin{figure*}[h!]
\centering
\includegraphics[width=0.90\textwidth]{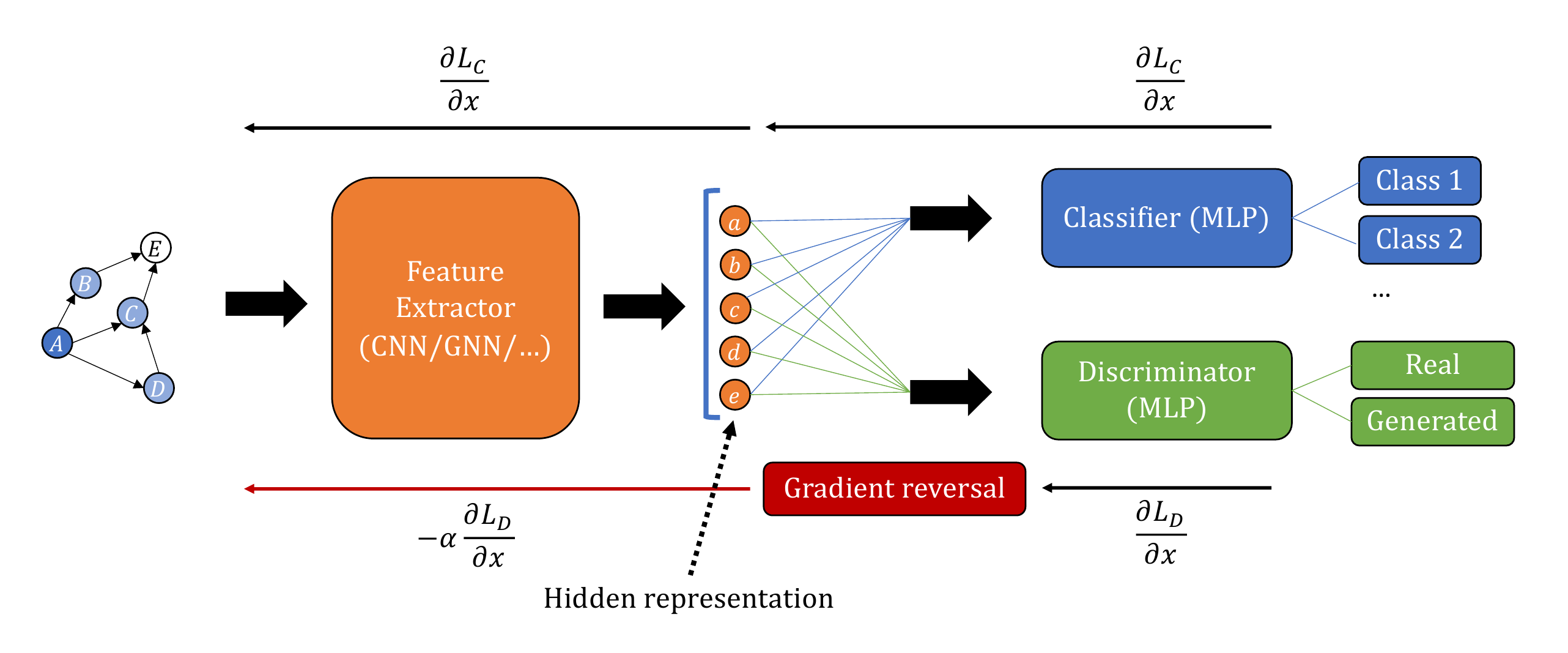}
\caption{Domain-adversarial network optimization process}
\label{fig:dagnn}
\end{figure*}

In practice this means that there are three networks: a base network which produces a latent space representation of the input and then two adversarial head networks.  The first head network, the discriminator, takes the latent space representation from the base network and makes a prediction whether the input came from the labelled training dataset or the unlabelled test dataset.  Then the loss is computed between those predictions and the actual domains from which the inputs came.  The discriminator loss $L_{D}$ is back-propagated through the discriminator to update its weights.  The second head network, the classifier, then takes the latent space representation from the base network and makes a prediction for our end classification task.  The classifier loss $L_C$ is back-propagated through the classifier, updating its weights.  Finally, the loss of the discriminator and the classifier are back-propagated through the base network with a \textbf{reversed} gradient from the discriminator

\begin{equation}
\label{loss}
\frac{\partial L_{total}}{\partial x} = -\alpha\frac{\partial L_{D}}{\partial x} +\frac{\partial L_{C}}{\partial x}.
\end{equation}
The reversed gradient from the discriminator loss acts as a penalizing term for the base network so that it is biased toward representations which the discriminator is unable to distinguish.  Here, $\alpha$ is another hyperparameter of the training, a coefficient that can depend on the training epoch or iteration.  After training, the discriminator is discarded and the network consists solely of the base network and the classifier.  The domain-adversarial process is depicted in figure~\ref{fig:dagnn}.

\subsection{Network Training and Evaluation} \label{Network Training and Evaluation}
Our training sample consisted of 140k events from the MC outbending sample filtered so that they each contained an identified $p$, $\pi^{-}$, and scattered $e^{-}$ in the reconstructed event.  The sample consisted of a random sample of half MC truth $\Lambda$ signal events and half MC truth background.  Signal events were identified as those events with a $\Lambda \rightarrow p\pi^{-}$ decay in the MC truth, and all other events were marked as background.  The real data sample for the domain-adversarial approach consisted of roughly 140k events from the fall 2018 outbending runs.  The samples were split $80\%$ for training, $10\%$ for model validation, and $10\%$ for model testing.

\subsubsection{Data Preprocessing} \label{Data Preprocessing}
The same preprocessing routine was applied to both MC and real data.  Events were mapped into fully connected graphs containing all the particles from the reconstructed event.  Each node in the graph represented a particle and its data consisted of the reconstructed transverse momentum $p_T$, polar and azimuthal angles $\theta$ and $\phi$, velocity $\beta$, PID, $\chi^2$ estimate for the PID assignment, and detector status.  The continuous quantities $p_T$, $\theta$, $\phi$, and $\beta$ first had the event mean for that variable subtracted from them and were then normalized to the maximum difference from the event mean.  The large default values for unassigned PID $\chi^2$ estimates were set to $+10$ since these sometimes correspond to hadrons for which the $\chi^2$ estimate is not required.  Then, the PID $\chi^2$ estimate was cut to be within the range $[-10,10]$ and normalized.  Discrete quantities, particle PID and detector status, were reassigned to arbitrary float values from $-1$ to $1$.  The PID reassignment values are listed in table~\ref{table:pid_reassignment}.  For charged particles the replacement value had the same sign as the charge of the particle and particle-antiparticle pairs had identical magnitudes.  Otherwise replacement value choice was arbitrary.  Detector status values are confined within $(-5000,5000)$ so the status was just normalized by $5000$.  The renormalization of inputs to values in $[-1,1]$ is a standard step in the machine learning process to avoid large gradient values during optimization.

\begin{table}[h!]
\centering
\begin{tabular}{ c | c }
 \hline\hline
True PID & Replacement value \\ 
\hline
 $22$      & $0.0$ \\
 $\pm11$   & $\mp1.0$ \\ 
 $\pm2212$ & $\pm0.8$ \\
 $2112$    & $0.5$ \\
 $111$     & $0.1$ \\
 $\pm211$  & $\pm0.6$ \\
 $311$     & $0.3$ \\
 $\pm321$  & $\pm0.4$ \\
 \hline
\end{tabular}
\label{table:pid_reassignment}
 \caption{PID replacement values used to renormalize discrete PID values.}
\end{table}

\subsubsection{Model Architecture} \label{Model Architecture}
The default architecture for the GIN consists of a $5$ layer network with each MLP having $2$ layers and a hidden dimension $hdim=64$.  The default final dropout rate is $0.5$ and the default graph and node pooling functions are a simple mean.  These architecture parameters (except pooling functions) were varied for the hyperparameter optimization search.  Each MLP layer used a ReLU activation function was followed with a batch normalization layer to reduce overfitting.  The final classification layer obviously only had $hdim=2$ corresponding to our two classification categories signal and background and was followed with a sigmoid activation function. Final classification assignments were normalized to probabilities with the softmax function

\begin{equation} \label{eq:softmax}
    \sigma(x) = \frac{e^{x_i}}{\sum_i e^{x_i}}.
\end{equation}

For the Domain-Adversarial GIN (DAGIN) the GIN architecture was the same except the dimension of the final layer still had the same hidden dimension as the previous layers.  The classifier and discriminator had identical architectures: each was an MLP with the same hidden dimension as the GIN in each layer except the final classification layer.  Each layer used ReLU activation except the final layer which used sigmoid activation.  Final classification predictions were also normalized with the softmax function from Eq.~\ref{eq:softmax}.  The hyperparameters left for optimization for the classifier and discriminator architecture were just the number of layers in each network.

\subsubsection{Training Parameters} \label{Training Parameters}
The ADAM optimizer~\cite{kingma2017adam} was used to train each model but the choice of batch size and learning rate $\eta$ was left for the hyperparameter optimization process.  The loss function for each model was a cross-entropy loss, except the discriminator which used binary cross-entropy loss.  For the DAGIN the learning rates of the three different networks base, classifier, and discriminator ($\eta$, $\eta_C$, $\eta_D$) were left as separate hyperparameters for optimization.

\subsubsection{Hyperparameter Optimization} \label{Hyperparameter Optimization}
A Bayesian optimization search of the hyperparameter phase space was carried out using the Optuna Hyperparameter Optimization package~\cite{optuna_2019}.  Optuna's builtin \textit{TPESampler}  was used which is based on the Tree-structured Parzen Estimator (TPE) algorithm~\cite{NIPS2011_86e8f7ab}.  The ROC (Receiver Operating Characteristic) curve is a plot of the fraction of true positives versus the fraction of false positives as a function of the threshold value for a binary classification task.  Hyperparameter choice was optimized based on the Area Under the (ROC) Curve (AUC) metric of trained model when evaluated on the reserved test data.  The more area under this curve, i.e. the higher the AUC, the less dependence there is on the threshold value and the more robust the model is.  This metric is preferred instead of a simple accuracy since it allows one to simultaneously maximize efficiency and purity for a binary classification task.

\begin{figure}[h!]
\centering
\includegraphics[width=0.45\textwidth]{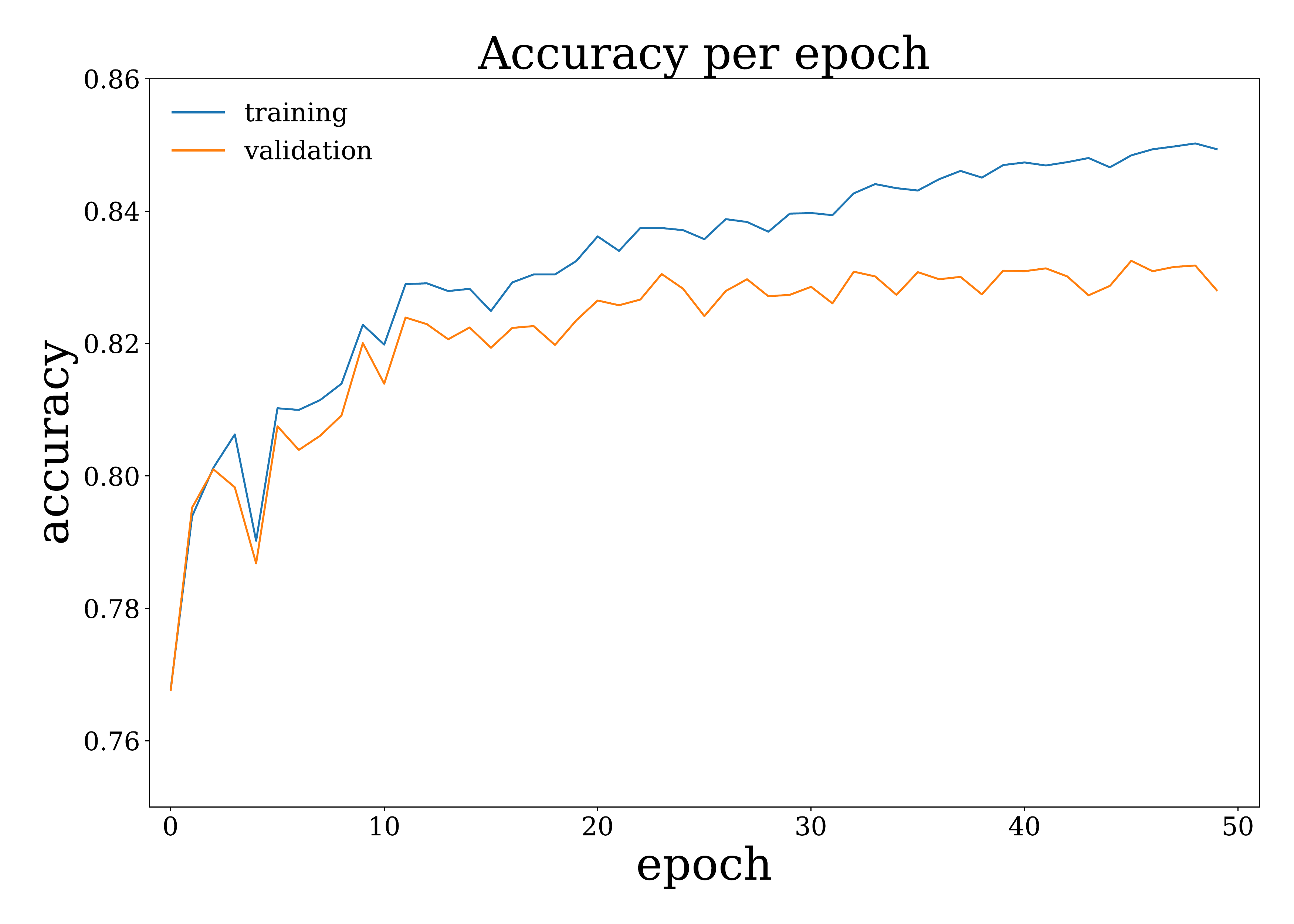}
\includegraphics[width=0.45\textwidth]{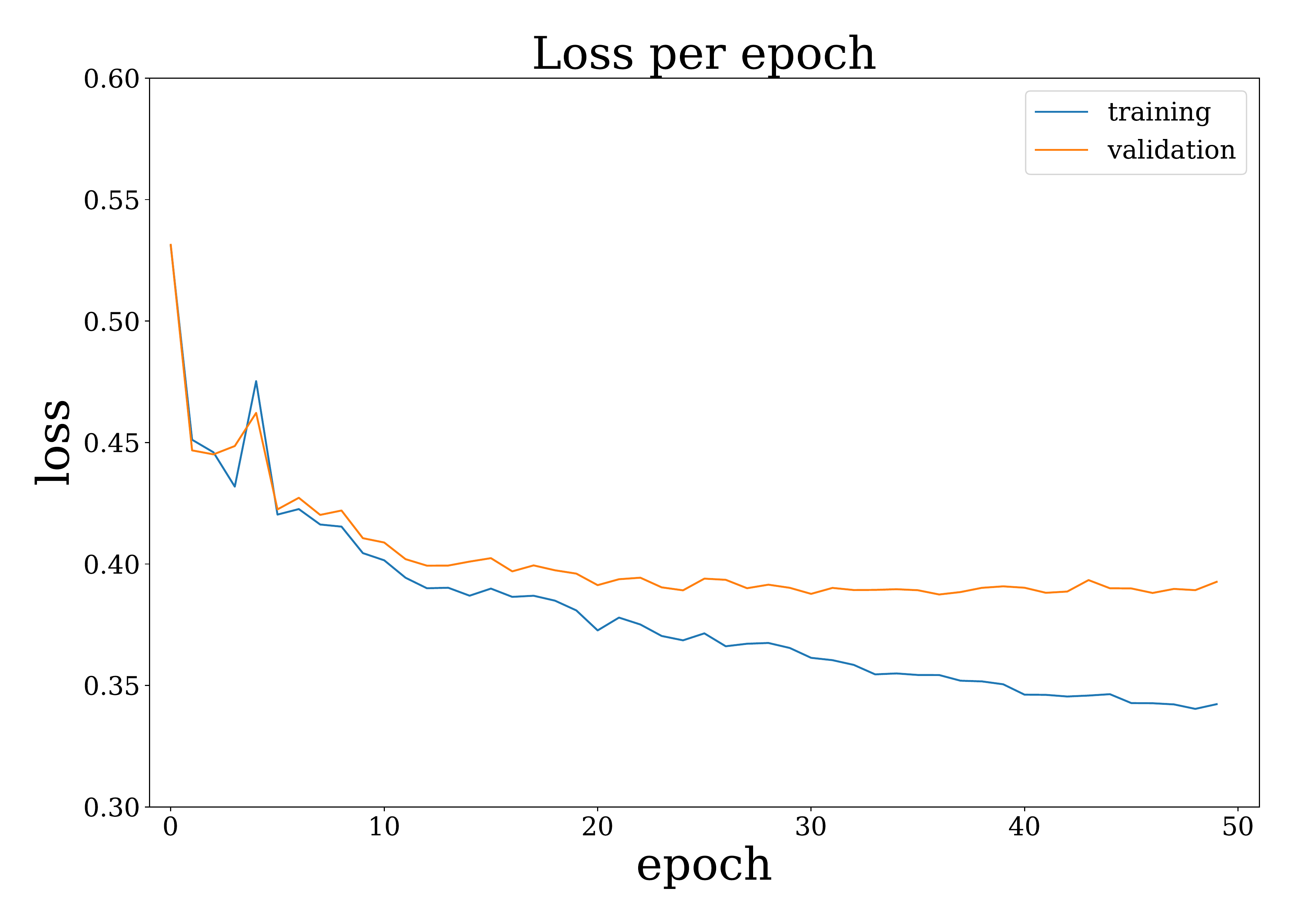}
\caption{Training accuracy (left) and loss (right) vs. epoch for the hyperparameter-optimized GIN.}
\label{fig:training_gin}
\end{figure}

\begin{figure}[h!]
\centering
\includegraphics[width=0.45\textwidth]{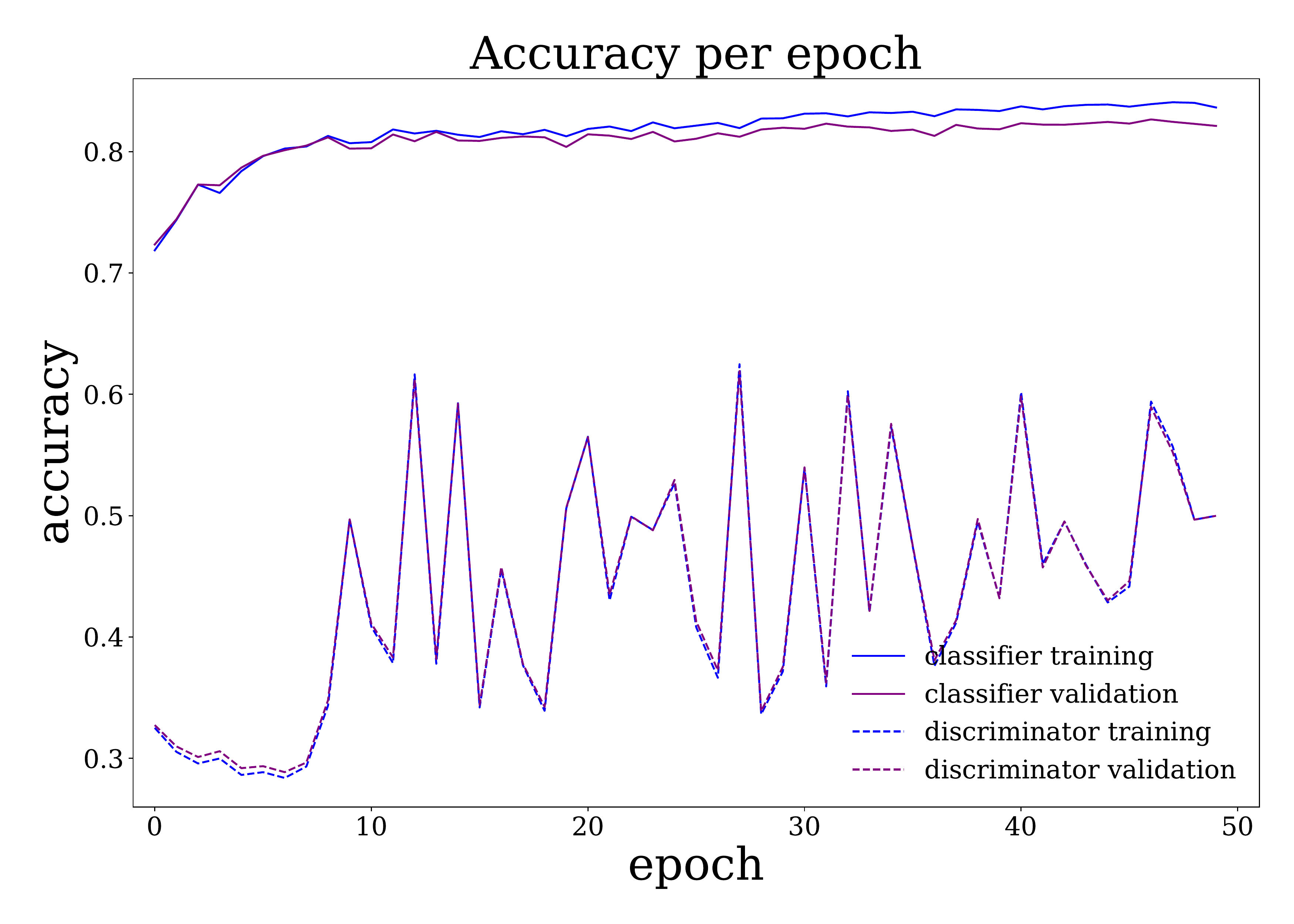}
\includegraphics[width=0.45\textwidth]{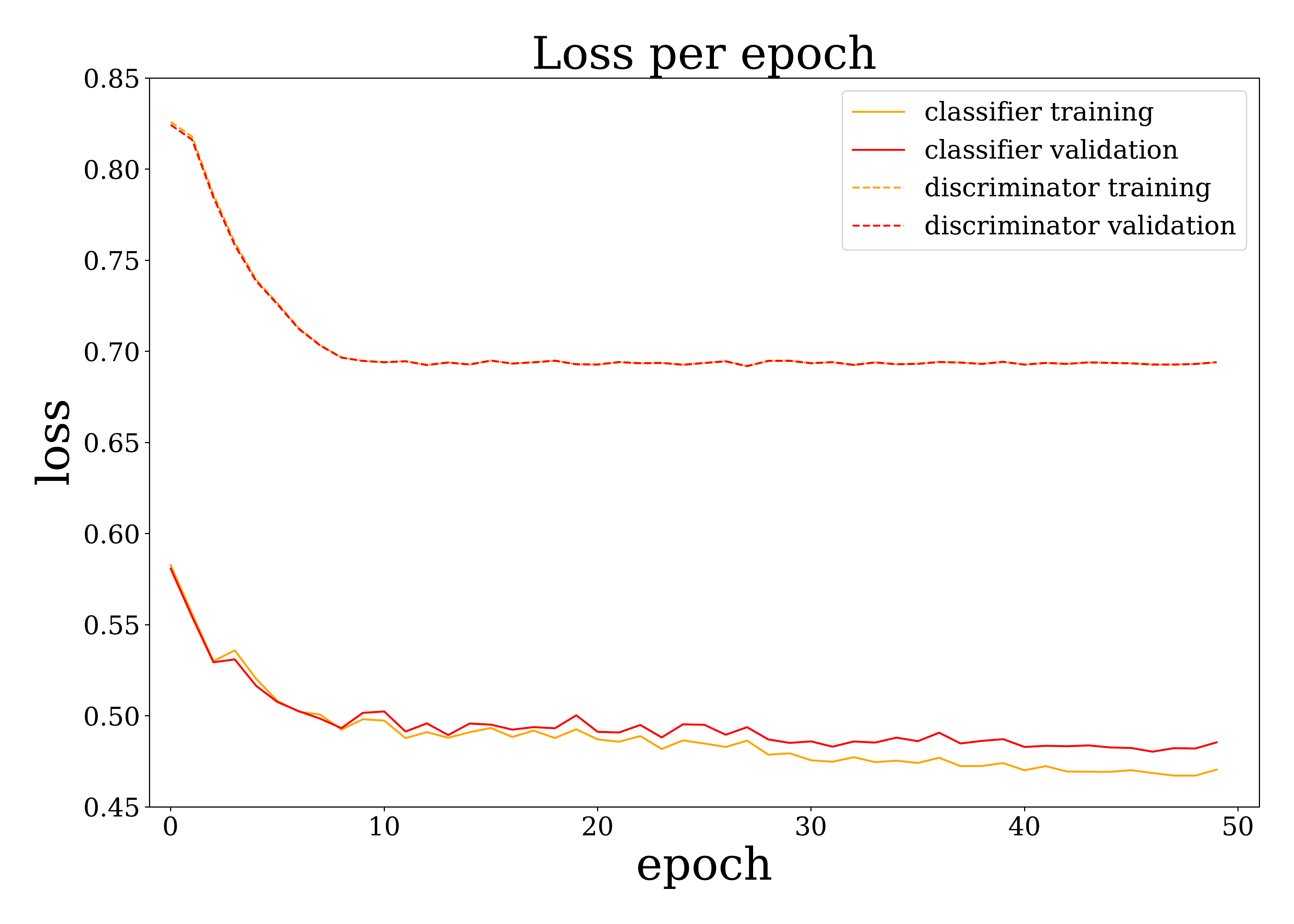}
\caption{Training accuracy (left) and loss (right) vs. epoch for the hyperparameter-optimized DAGIN.}
\label{fig:training_dagin}
\end{figure}

The hyperparameter phase space allowed for the optimization search is listed for the basic GIN model in table~\ref{table:hyperparameters_basic_GIN} and for the DAGIN in table~\ref{table:hyperparameters_DAGIN}.  Hyperparameters with ranges denoted in powers of $2$ or $10$ were sampled with a uniform log distribution and all other parameters were sampled with a uniform distribution.  Roughly $1500$ trials were performed for each hyperparameter search and each trial model was trained for up to $50$ epochs.  The training and validation metrics as a function of training epoch for the GIN and DAGIN are shown in Figs.~\ref{fig:training_gin} and \ref{fig:training_dagin} respectively.

\begin{table}[h!]
\centering
\begin{tabular}{ c | c }
 \hline\hline
 Hyperparameter & Range \\ 
 \hline
 nlayers & $[2,8]$ \\ 
 hdim & $[2^5,2^8]$ \\
 MLP nlayers & $[2,4]$\\
 dropout & $[0.5,0.8]$ \\
 batch & $[2^{6},2^{8}]$ \\
 $\eta$ & $[10^{-5},10^{-2}]$\\ 
 \hline
\end{tabular}
\label{table:hyperparameters_basic_GIN}
\caption{Hyperparameter ranges for basic GIN optimization search.}
\end{table}

\begin{table}[h!]
\centering
\begin{tabular}{ c | c }
\hline\hline
Hyperparameter & Range \\ 
\hline
 nlayers & $[2,8]$ \\ 
 Classifier/Discriminator nlayers & $[2,8]$ \\
 hdim & $[2^5,2^8]$ \\
 Classifier/Discriminator hdim & $[2^5,2^8]$ \\
 MLP nlayers & $[2,4]$\\
 dropout & $[0.5,0.8]$ \\
 batch & $[2^{6},2^{8}]$ \\
 $\eta$ & $[10^{-5},10^{-2}]$\\ 
 $\eta_C$ & $[10^{-5},10^{-2}]$\\
 $\eta_D$ & $[10^{-5},10^{-2}]$\\
 $\alpha$ & $[1,100]$ \\
 \hline
\end{tabular}
\label{table:hyperparameters_DAGIN}
 \caption{Hyperparameter ranges for DAGIN optimization search. $\alpha$ is the coefficient for the discriminator loss in the gradient-reversal layer.}
\end{table}

As a sanity check, this same hyperparameter optimization routine was run with similar datasets filtered so that the MC background events came from varying fractions of the two main background contributions: combinatorial background and $\Delta \rightarrow p\pi^{-}$ decays.  The performance achieved with these datasets was essentially the same as that achieved on the original dataset, demonstrating that there was not any wild dependence of the training process on the composition of the background sample.

\section{Results} \label{Results}

\subsection{Evaluation on Monte Carlo} \label{Evaluation on Monte Carlo}
\begin{figure}[h!]
\centering
\includegraphics[width=0.45\textwidth]{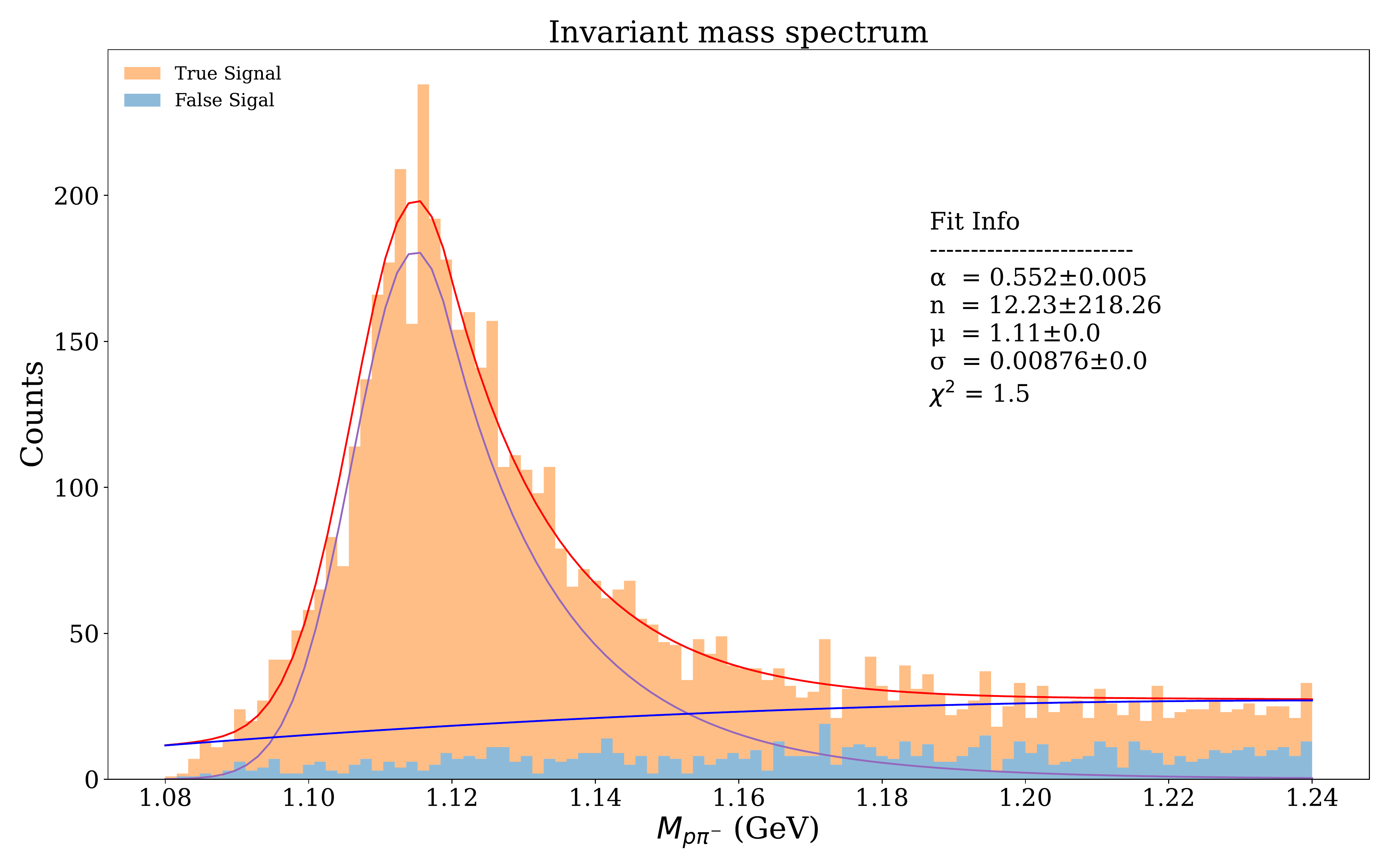}
\includegraphics[width=0.45\textwidth]{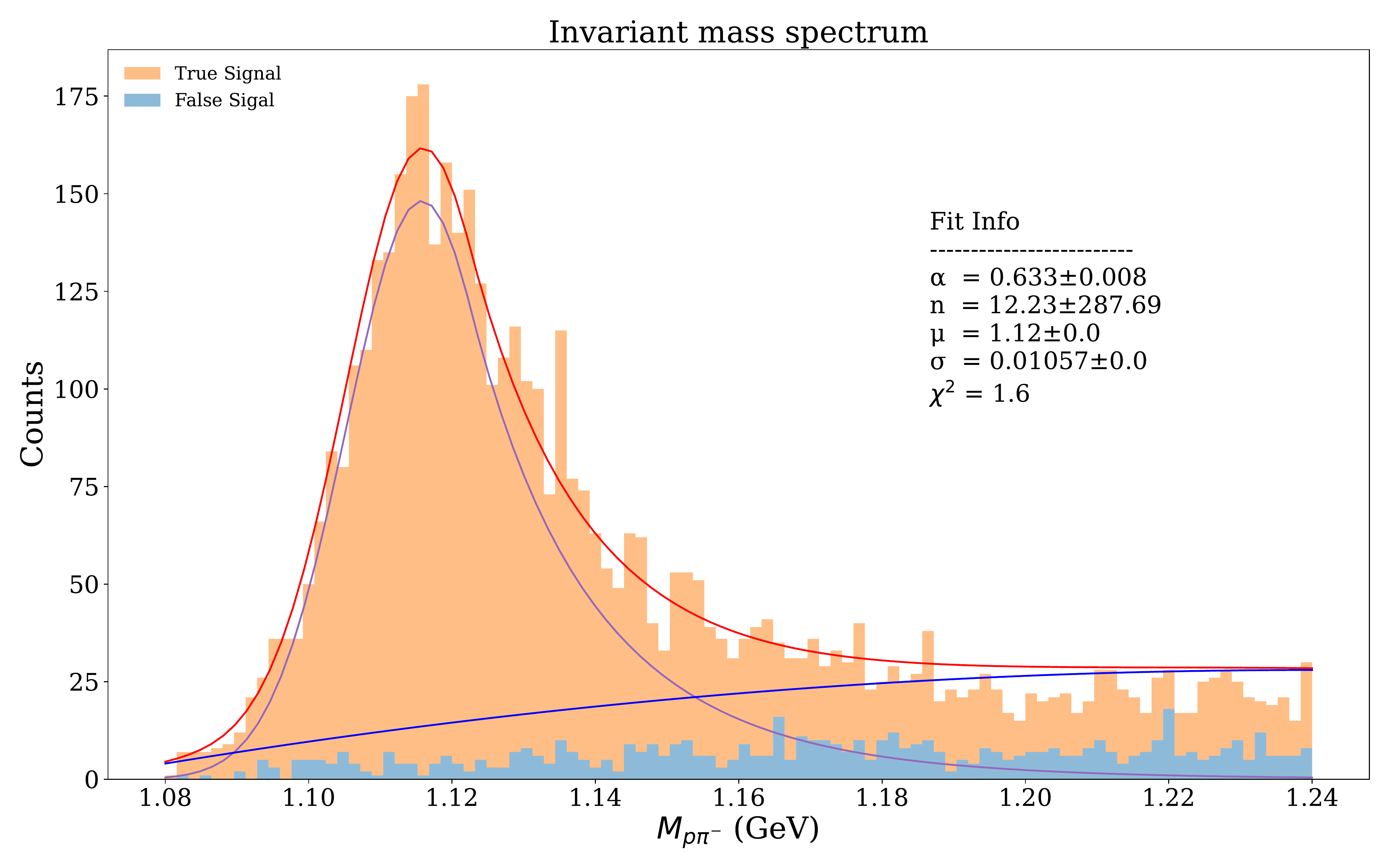}
\caption{Signal mass spectra identified by the hyperparameter-optimized GIN (left) and DAGIN (right) with signal extraction fits applied.  The stacked histograms are separated into true signal (orange) and false signal (blue).}
\label{fig:MC_evaluation}
\end{figure}

\begin{figure}[h!]
\centering
\includegraphics[width=0.45\textwidth]{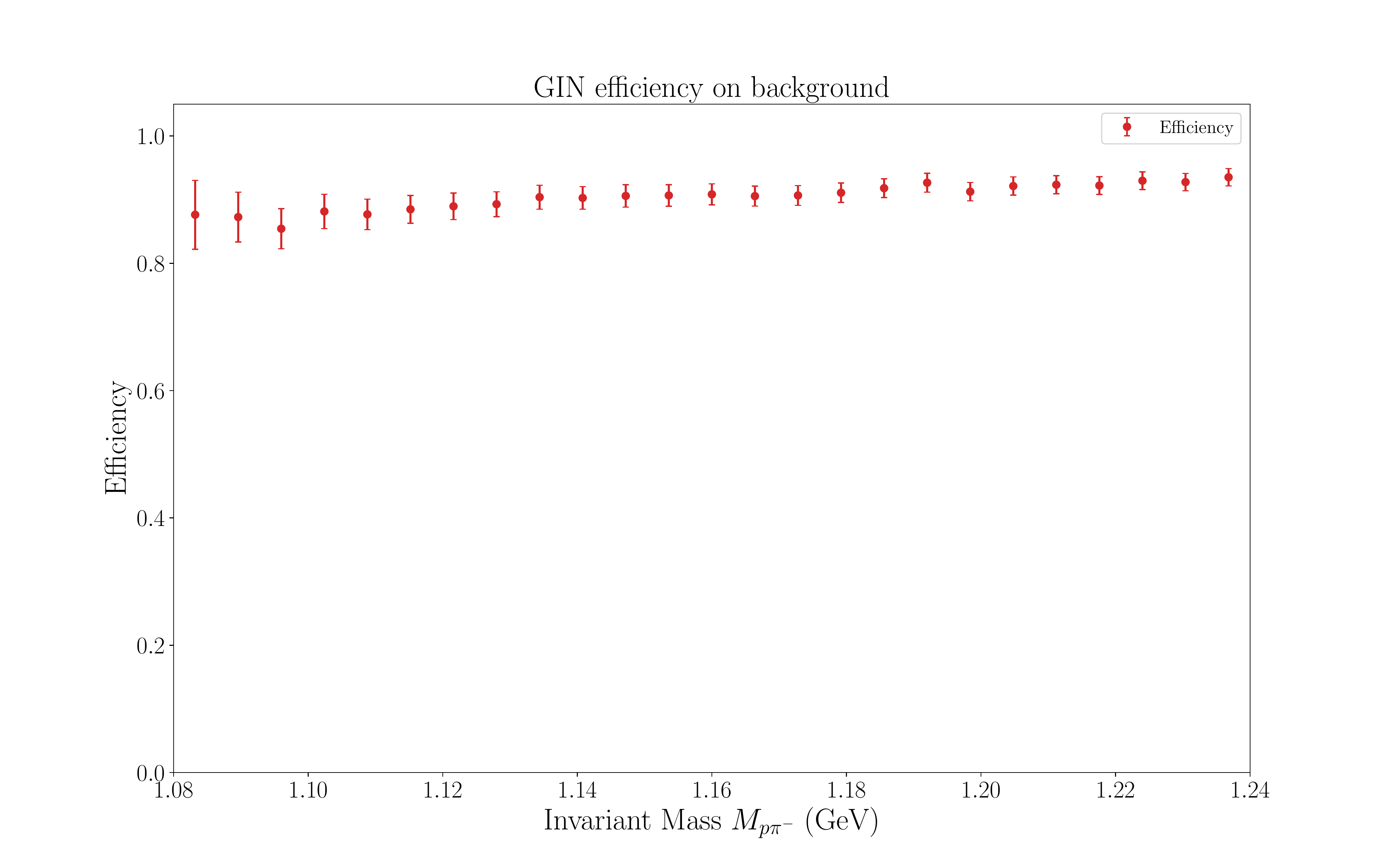}
\includegraphics[width=0.45\textwidth]{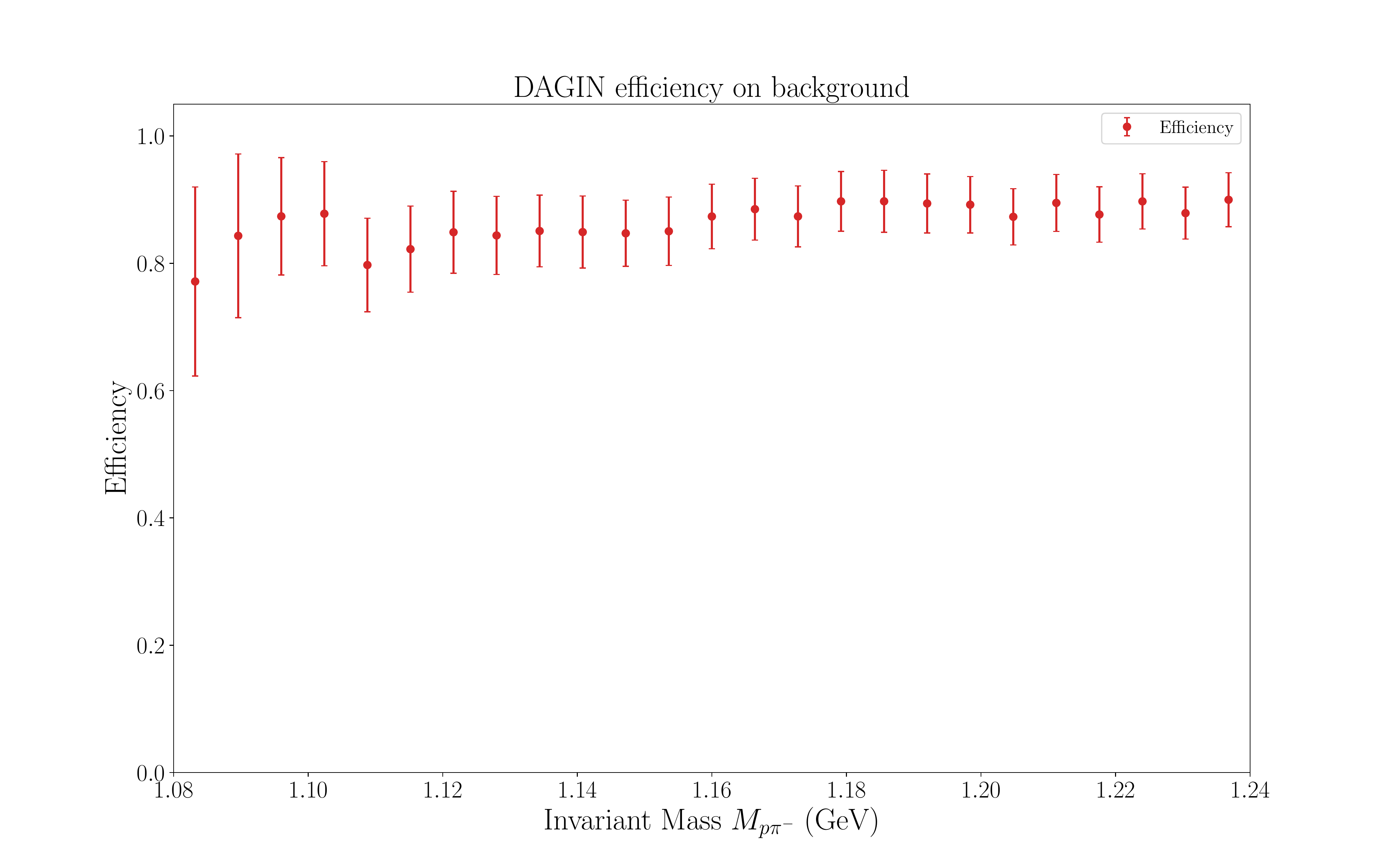}
\caption{Efficiency of the GIN (left) and DAGIN (right) using the \textit{MoDeLoss} decorrelation method on background events as a function of the invariant mass.  Errors are solely statistical.}
\label{fig:efficiency}
\end{figure}

After training, the performance of each model was evaluated on the reserved $10\%$ of the MC dataset.  The invariant mass spectrum of the GNN-identified signal events was fit to the same Crystal Ball signal over quadratic background function to determine if the signal shape was changed at all by the application of the GNN.  Note that two of the fit parameters for the background quadratic fit function, those describing the peak height and location of the parabola, were held constant and only the curvature parameter of the background was varied for the fit.  Both the GIN and DAGIN roughly preserve the signal shape as shown in figure~\ref{fig:MC_evaluation} for the optimized models.  There is some increase in the $\sigma$ parameter of the Crystal Ball fit function, particularly for the DAGIN.  However, this is mainly due to a loss in signal efficiency centered around the signal peak.  This eliminates the slight excess above the fit function at the signal peak evident in figure~\ref{fig:mass} allowing the $\sigma$ parameter to increase without actually significantly impacting the width of the signal.  The efficiency of the network on background events is fairly stable as a function of the invariant mass as shown in figure~\ref{fig:efficiency}.  Employing the decorrelation method described in ref.~\cite{Kitouni:2020xgb} as a sanity check to decorrelate the performance of the network from the invariant mass produced similar results.  Efficiencies of the GIN and DAGIN networks trained with the decorrelation method are shown in figure~\ref{fig:efficiency_modeloss}.  The details of our setup are described in Appendix~\ref{Appendix}.
 




\begin{figure}[h!]
\centering
\includegraphics[width=0.45\textwidth]{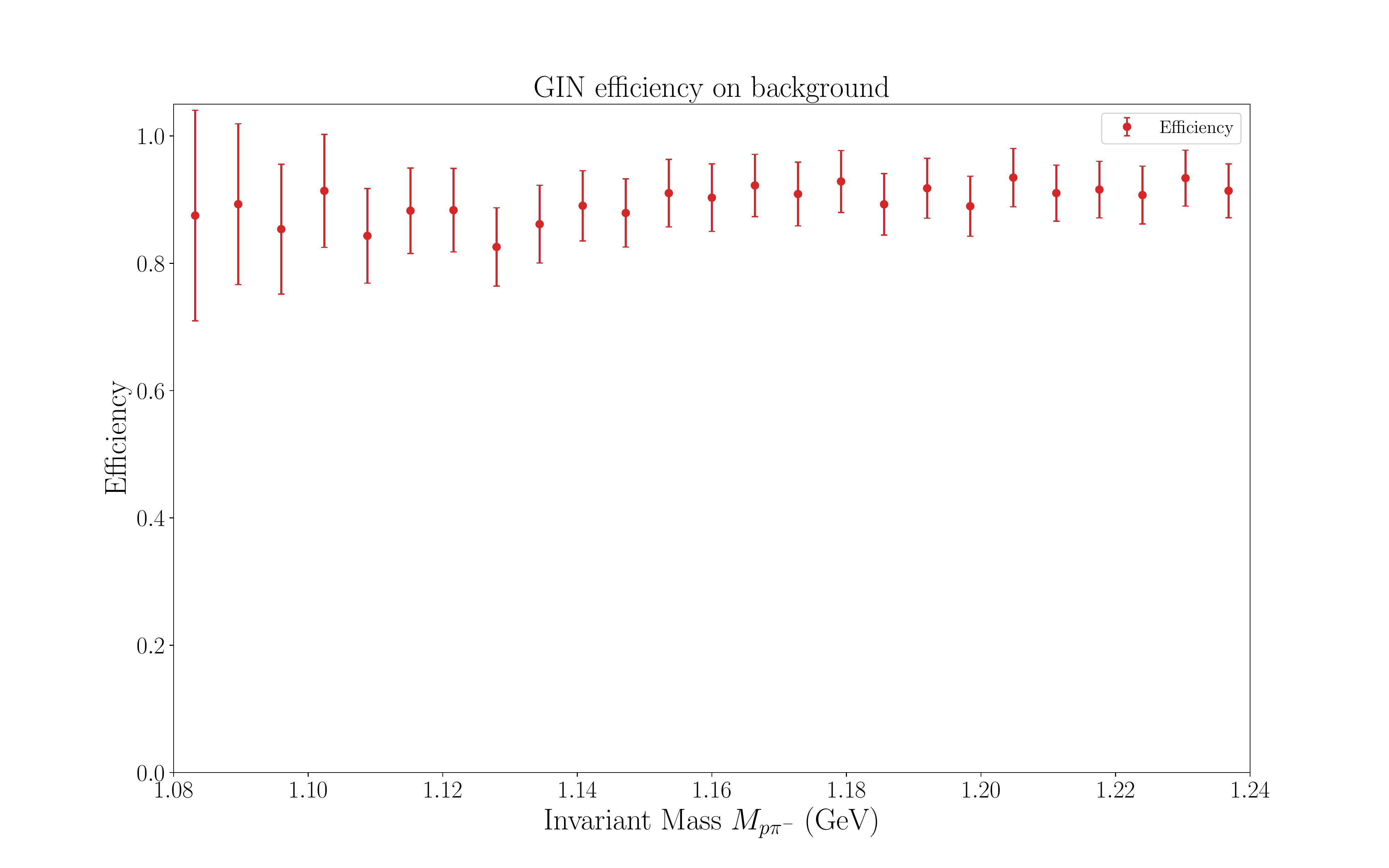}
\includegraphics[width=0.45\textwidth]{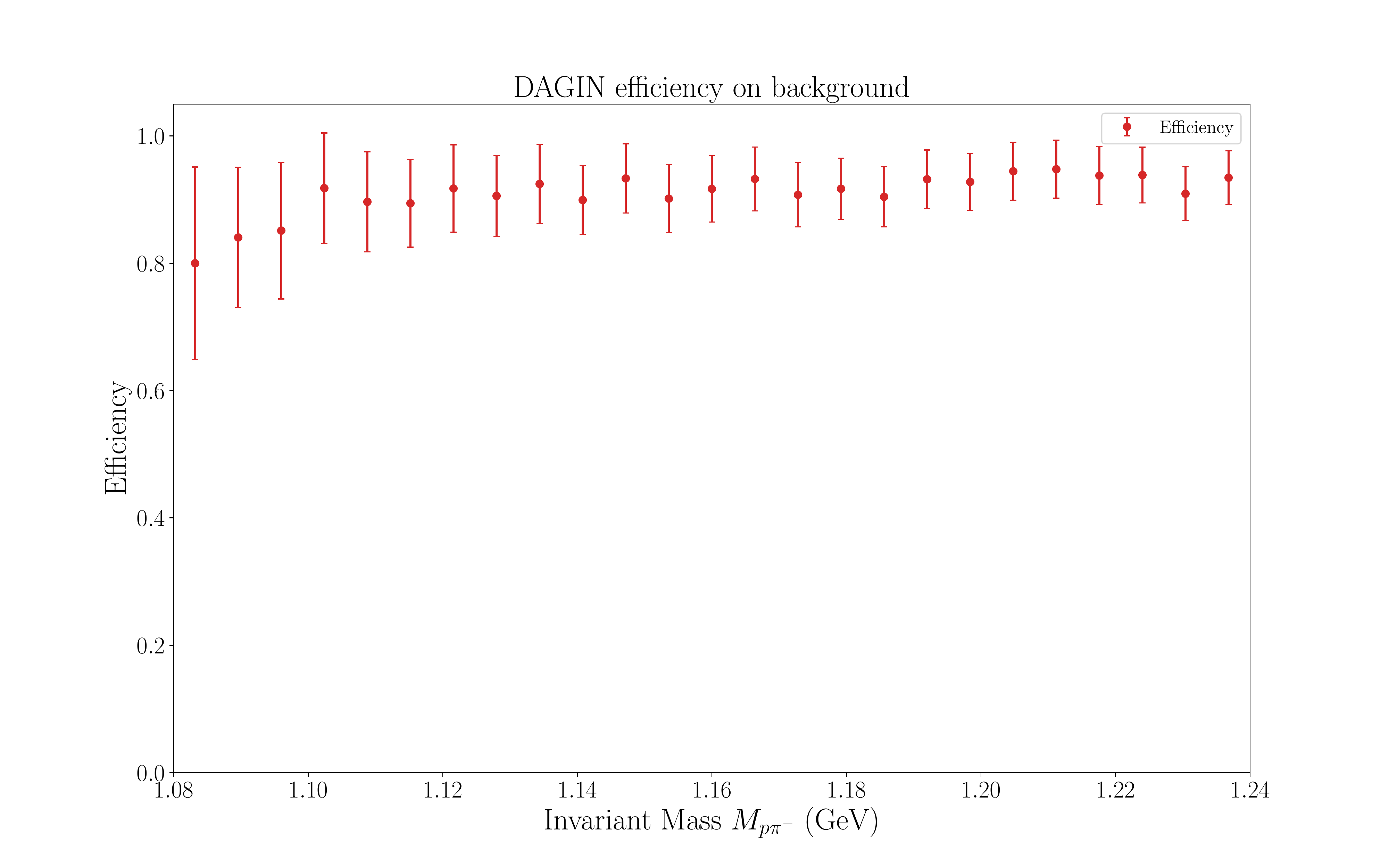}
\caption{Efficiency of the hyperparameter-optimized GIN (left) and DAGIN (right) on background events as a function of the invariant mass.  Errors are solely statistical.}
\label{fig:efficiency_modeloss} 
\end{figure}

\subsection{Evaluation on Data} \label{Evaluation on Data}
\begin{figure}[tbh]
\centering
\includegraphics[width=0.45\textwidth]{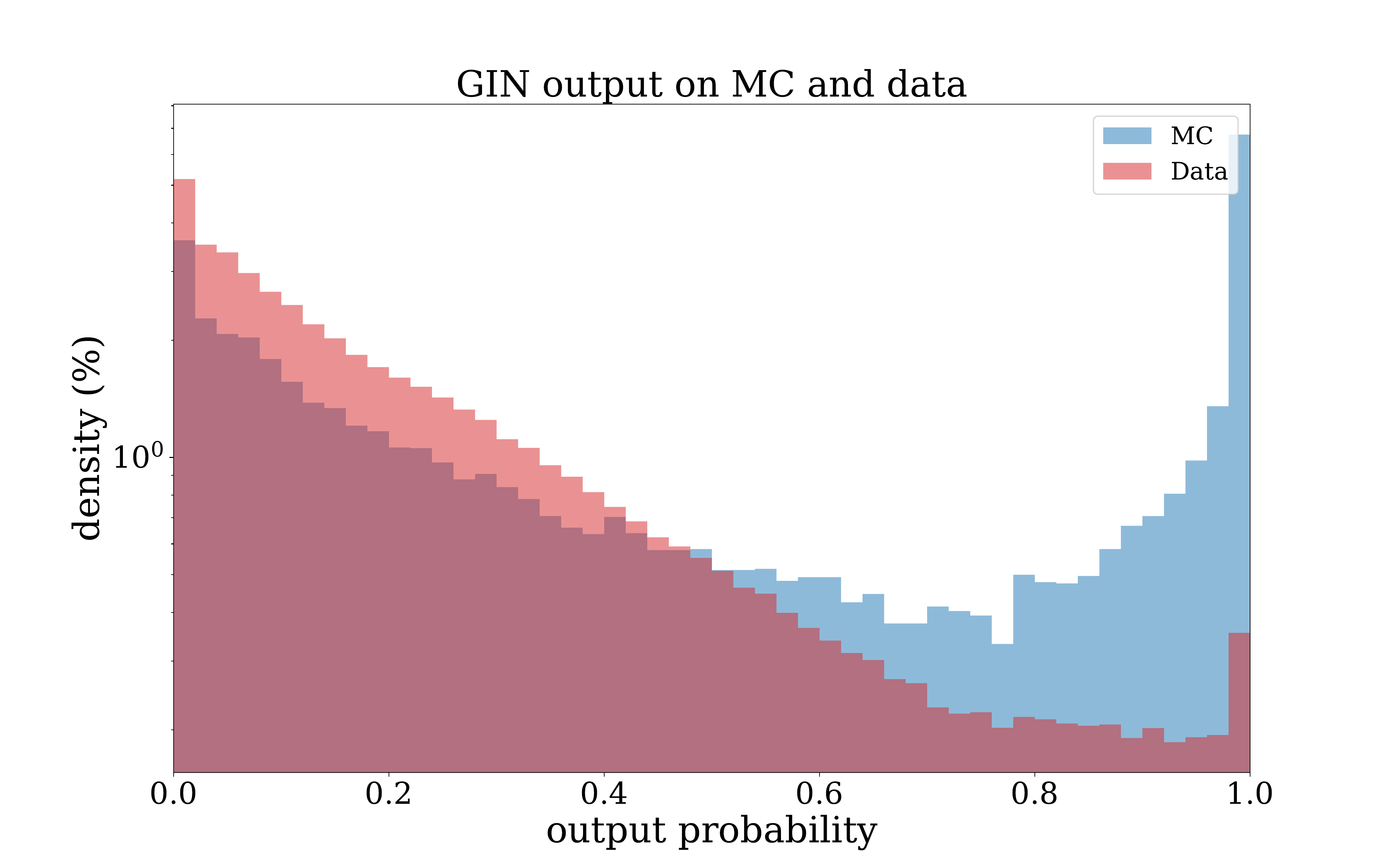}
\includegraphics[width=0.45\textwidth]{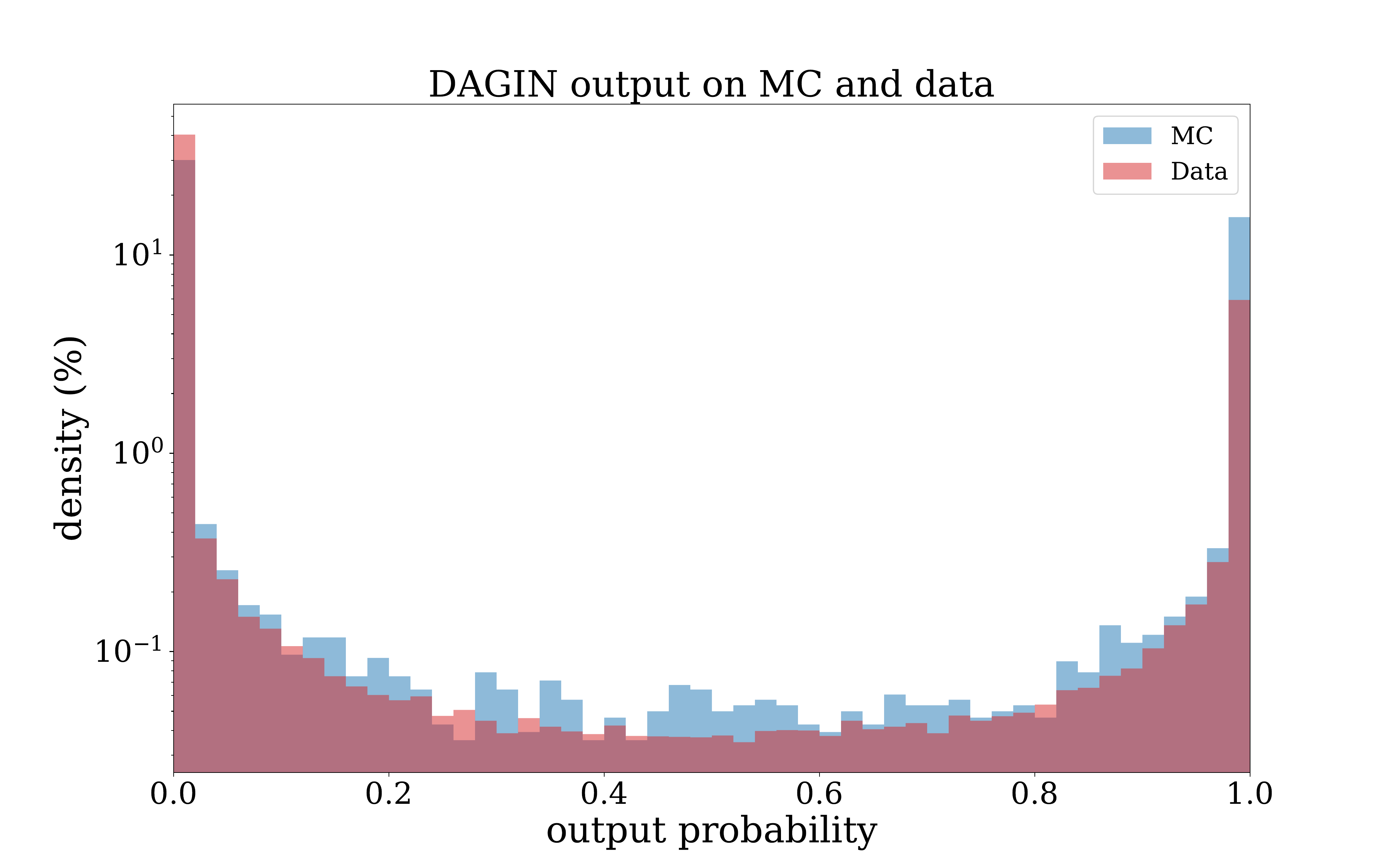}
\caption{Output distributions on real data and MC for the GIN (left) and DAGIN (right).  The NN output probability spectrum is restricted within $[0.0,1.0]$ where $0.0$ corresponds to a background event and $1.0$ corresponds to a $\Lambda$ signal event.}
\label{fig:nn_outputs}
\end{figure}

\begin{figure}[tbh]
\centering
\includegraphics[width=0.45\textwidth]{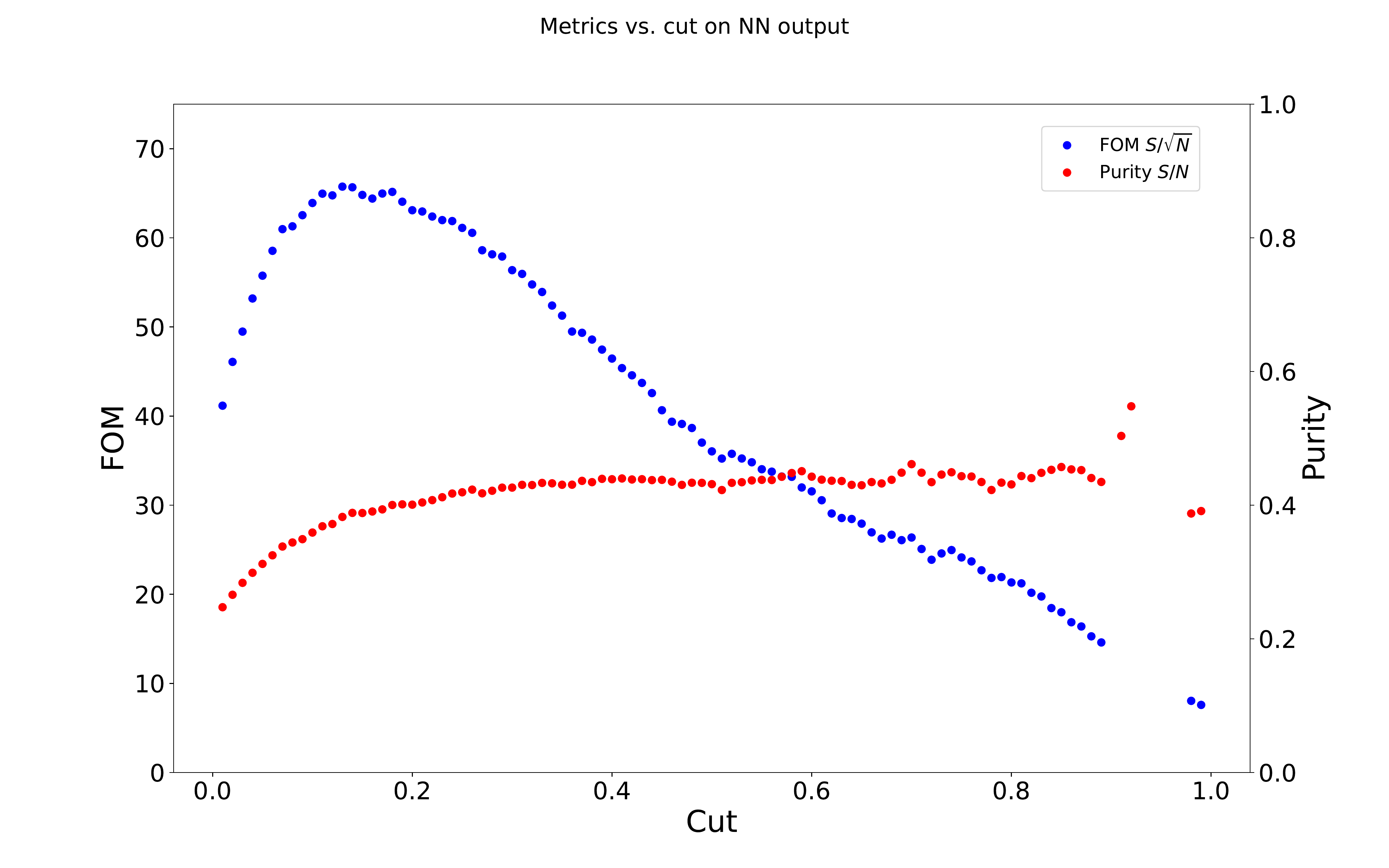}
\includegraphics[width=0.45\textwidth]{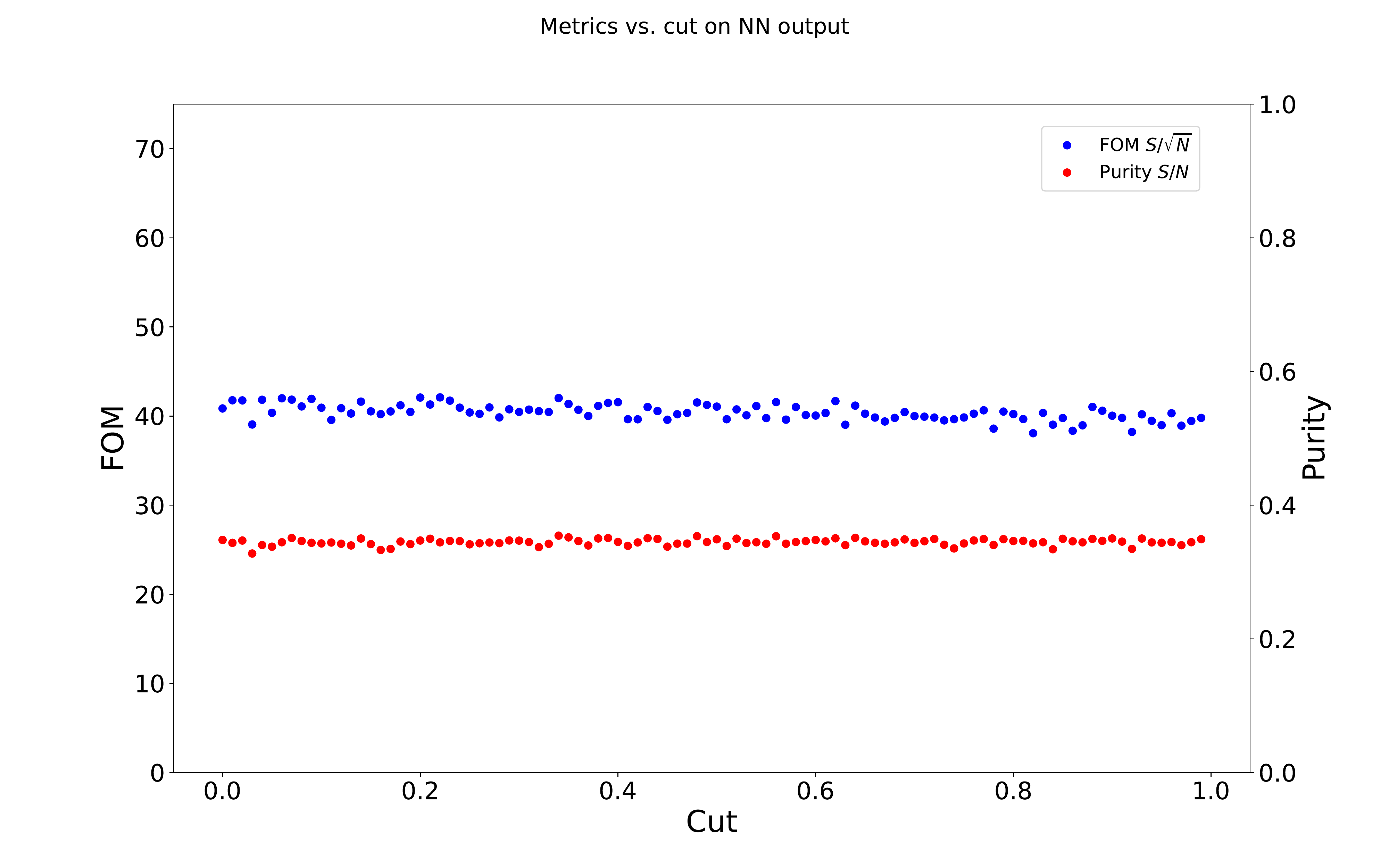} 
\caption{Scans of the FOM (blue) and purity (red) from the signal over background fits on data as a function of the cut on the GNN output for the GIN (left) and DAGIN (right).  All events with GNN output greater than the cut value are identified as signal events.}
\label{fig:roc_scan}
\end{figure}

\begin{figure}[tbh]
\centering
\includegraphics[width=0.45\textwidth]{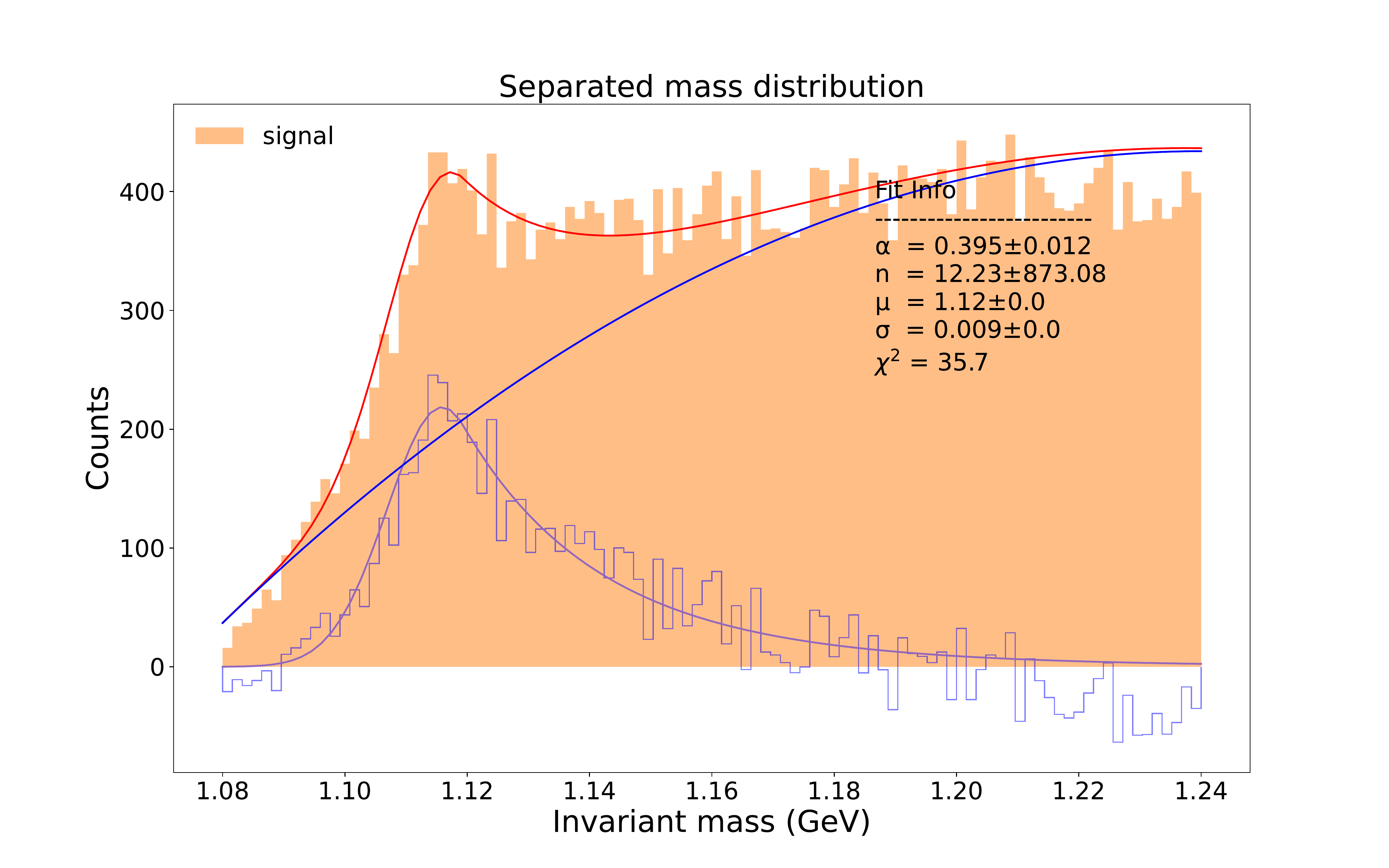}
\includegraphics[width=0.45\textwidth]{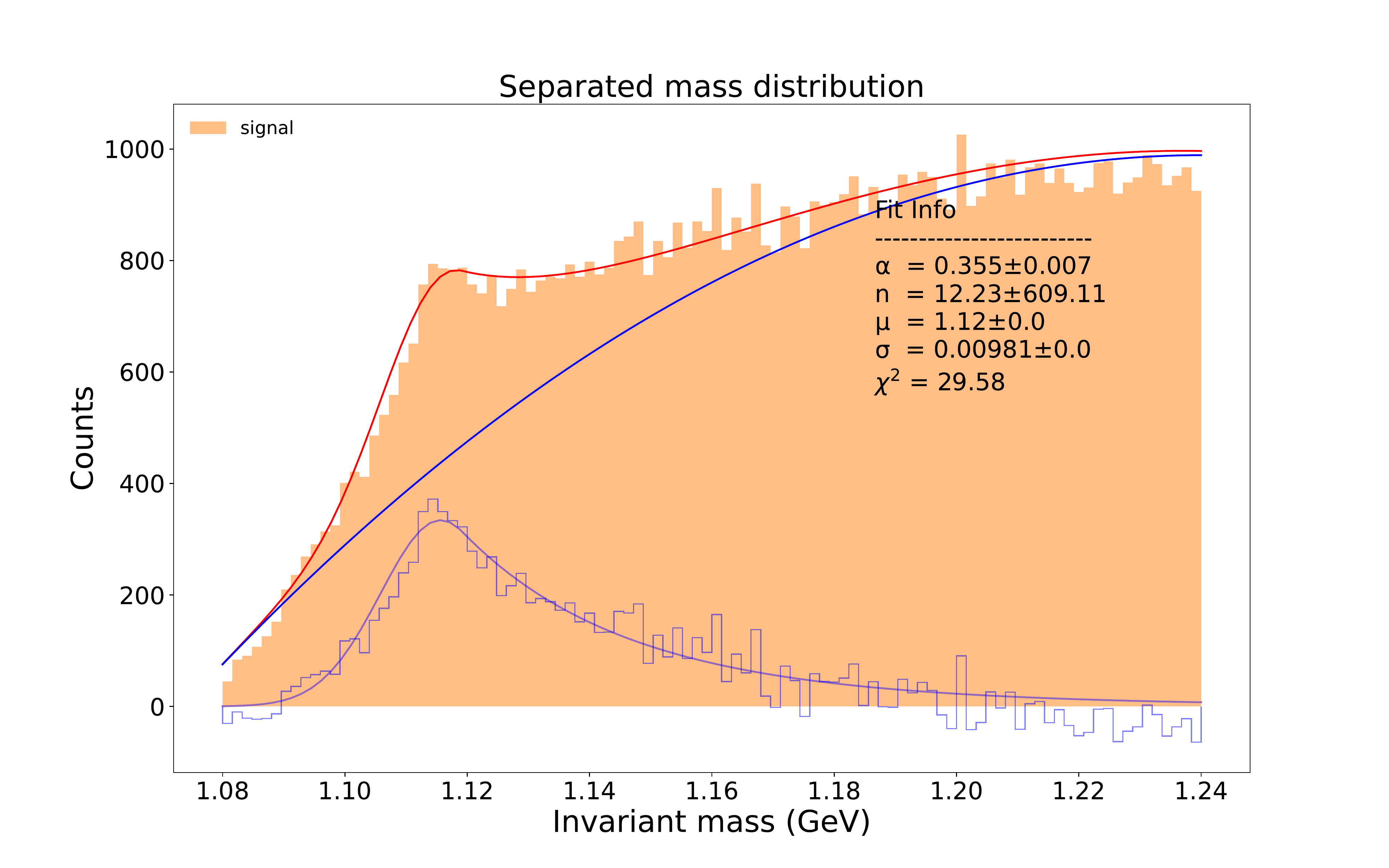}
\caption{The same Crystal Ball signal over quadratic background fit was applied to the NN-identified signal mass spectrum (pale orange histogram) for the GIN (left) and DAGIN (right).  $N_{sig}$ is estimated using the histogram counts over the fitted background function (small outlined histogram).}
\label{fig:data_evaluation}
\end{figure}

The figure of merit ($FOM=N_{sig}/\sqrt{N_{tot}}$) used to evaluate the performance on data is calculated within the $\pm2\sigma$ region around the signal peak location $\mu$ from the fit to the mass spectrum of events the GNN identifies as signal.  Maximizing this FOM should simultaneously maximize the purity $N_{sig}/N_{tot}$ while minimizing the error, $1/\sqrt{N_{tot}}$ assuming Poissonian statistics.  The FOM and purity were computed for various values of the cut on the GNN output in order to maximize the FOM and purity.  The GNN output distributions on data are shown in comparison with the distributions on MC in figure~\ref{fig:nn_outputs}, and the scans of the cut values are shown in figure~\ref{fig:roc_scan} for both GIN and DAGIN.  Note that the output probability from the DAGIN is taken just after the final classification layer without applying the softmax function.  The maximal FOMs and corresponding purities obtained are listed in table~\ref{table:FOM}.  The invariant mass spectra after the application of the optimized GIN and DAGIN respectively with maximized FOM are shown in figure~\ref{fig:data_evaluation}.  The signal shape is still relatively well preserved by the application of either the GIN or DAGIN.  The Kolmogorov-Smirnov distances between the NN output distributions and data and MC are also computed.  These values are $0.262$ and $0.217$ for the GIN and DAGIN respectively. 

\begin{table}[h!]
\centering
\begin{tabular}{ c | c c c }
\hline\hline
Method & No GNN & GIN & DAGIN \\
\hline
$N_{sig}/\sqrt{N_{tot}}$ & 34.11 & 65.74 & 42.09 \\ 
$N_{sig}/N_{tot}$        & 0.195 & 0.382 & 0.354 \\ 
NN cut                   &  N/A  & 0.13  & 0.22  \\ 
\hline
\end{tabular}
\label{table:FOM}
 \caption{Figure of Merit from evaluation on data for all three methods.}
\end{table}

\section{Conclusions} \label{Conclusions} 
We have developed a GNN architecture for tagging $\Lambda$ hyperon events that reaches up to $83\%$ test accuracy and $0.9$ AUC values after training.  The method also roughly preserves the $\Lambda$ signal shape in both data and simulation.  While there remains room for further improvement, already the GIN model increases the $\Lambda$ signal purity by a factor of $1.95$.  Interestingly, the DAGIN model only increases the purity by a factor of $1.82$.  This could occur if the features that the domain-adversarial method forces the network to use are not as discriminating as those used by the GIN even if they are common to both datasets.  However, the DAGIN model produces a more consistent output between data and MC with a Kolmogorov-Smirnov statistic of $0.217$ compared to the GIN with a statistic of $0.262$.  The methods developed here may also be applied to other channels at CLAS12 and provide a baseline for future efforts using GNNs for event tagging in lower energy regimes.  Another potential use case could be extending these networks to filter out feed-down contributions to the $\Lambda$ signal from decays of heavier hyperons, e.g. $\Sigma^{0} \rightarrow \Lambda \gamma$.  This would be especially useful at the planned Electron Ion Collider where feed-down is expected to be more of an issue~\cite{Kang_2022}.  We plan to further check this method by evaluating how much it improves $\Lambda$ longitudinal spin transfer measurements at CLAS12.  Future improvements include exploring other network architectures, specifically Subgraph Networks, which are theoretically even more powerful than GINs~\cite{https://doi.org/10.48550/arxiv.2206.11140} and expanding the input data to the graphs to include detector level information.

\section*{Acknowledgements}
We acknowledge the outstanding efforts of the staff of the Accelerator, the Physics Divisions at Jefferson Lab, and the CLAS Collaboration in making this experiment possible.

\paragraph{Funding information}
This material is supported by the U.S. Department of Energy, Office of
Science, Office of Nuclear Physics under Award Numbers DE-SC0019230 and DE-AC05-06OR23177. 

\appendix

\section{Decorrelation} \label{Appendix}
Ref.~\cite{Kitouni:2020xgb} introduces a method for training classifiers to be independent of or have a polynomial dependence on a mass parameter, which is important for resonance searches.  Other methods for mass decorrelation exist, including planing, adversaries, Distance Correlation, and Flatness~\cite{Kitouni:2020xgb}.  The following discussion gives a brief introduction of this method closely following ref.~\cite{Kitouni:2020xgb} and outlines our implementation.

Often in resonance searches it is necessary to enhance an invariant mass signal in such a way that the enhancement selection is decorrelated from the actual mass spectrum.  Neural network classifiers are powerful statistical methods which can often learn the invariant mass spectrum given event-level particle information leading to an event selection biased by the mass parameter of the resonance.  Ref.~\cite{Kitouni:2020xgb} introduced a penalizing loss term
\begin{equation}
    L_{MODE}^{\ell} = \sum_m |F_m(s) - F_m^{\ell}(s)|^2 ds
\end{equation}
where $s$ is the score output by the classifier and used for prediction while $F_m(s)$ is the Cumulative Distribution Function up to scores $s$ and $m$ is an index referring to the mass bin. $F^{\ell}_m(s)$ is defined by
\begin{equation}
    F_m^{\ell} = \sum_{l = 0}^{\ell} c_l(s) P_l(\tilde{m})
\end{equation}
where $c_{\ell}(s)$ is the $\ell^{th}$ Legendre moment and $P_{\ell}(\tilde{m})$ is the $\ell^{th}$ Legendre polynomial as a function of $\tilde{m}$ the central mass value of the $m^{th}$ mass bin.  This penalty term essentially forces a classifier to converge towards a feature space that does not have the specified $\ell$ polynomial dependence on the mass parameter of the resonance.  For our brief foray with this method, we used the invariant mass of the $p\pi^{-}$ pair as our mass parameter and required independence ($\ell=0$) from the mass.  The coefficient on the penalty term was set to $1$ so that the overall loss was \begin{equation}
    L_{Total} = L_{C} + L_{MODE}
\end{equation}
where $L_C$ is the base loss of the classifier.  We used the implementation of the loss function provided by the authors of Ref.\cite{Kitouni:2020xgb} in their Python package \href{https://pypi.org/project/modeloss/}{modeloss}.  All other training details were kept the same.


\bibliographystyle{JHEP}
\bibliography{biblio.bib}

\end{document}